\documentstyle[psfig]{mn}

\def\etal{{\it et al.\ }}
\def\eg{{\it e.g.\ }}

\def\spose#1{\hbox to 0pt{#1\hss}}
\def\approxlt{\mathrel{\spose{\lower 3pt\hbox{$\sim$}}
	\raise 2.0pt\hbox{$<$}}}
\def\approxgt{\mathrel{\spose{\lower 3pt\hbox{$\sim$}}
	\raise 2.0pt\hbox{$>$}}}
\def\approxpropto{\mathrel{\spose{\lower 3pt\hbox{$\sim$}}
	\raise 2.0pt\hbox{$\propto$}}}
\mathchardef\twiddle="2218

\def\multleft#1{\hbox to size{\vbox {\halign {\lft{##}\cr #1}}\hfill}\par}
\def\multright#1{\hbox to size{\vbox {\halign {\rt{##}\cr #1}}\hfill}\par}

\def\today{\ifcase\month\or January\or February\or March\or April\or May\or
      June\or July\or August\or September\or October\or November\or December\fi
      \space\number\day, \number\year}
\def\<{\thinspace}

\def\apc{\rm atom cm$^{-2}$}

\def\cm{{\rm\thinspace cm}}
\def\erg{{\rm\thinspace erg}}

\def\K{{\rm\thinspace K}}

\def\km{{\rm\thinspace km}}
\def\kpc{{\rm\thinspace kpc}}

\def\Mpc{{\rm\thinspace Mpc}}
\def\Msun{\hbox{$\rm\thinspace M_{\odot}$}}

\def\s{{\rm\thinspace s}}
\def\yr{{\rm\thinspace yr}}

\def\ergpcmsqps{\hbox{$\erg\cm^{-2}\s^{-1}\,$}}

\def\ergps{\hbox{$\erg\s^{-1}\,$}}

\def\kmps{\hbox{$\km\s^{-1}\,$}}

\def\Msunpyr{\hbox{$\Msun\yr^{-1}\,$}}

\def\kmpspMpc{\hbox{$\kmps\Mpc^{-1}$}}

\def\apc{\rm atom cm$^{-2}$}
\def\pccmK{\hbox{$\cm^{-3}\K$}}
\def\pscm{\hbox{$\cm^{-2}\,$}}
\def\pccm{\hbox{$\cm^{-3}\,$}}

\title[Chandra X-ray observations of the 3C295 cluster core]
{Chandra X-ray observations of the 3C295 cluster core}  
\author[S.W. Allen et al.]
{\parbox[]{6.in} {S.W. Allen$^{1,4}$, G.B. Taylor$^{2}$, P.E.J. Nulsen$^{3,4}$,
R.M. Johnstone$^{1}$, L.P. David$^{4}$, S. Ettori$^{1}$, A.C. Fabian$^{1}$, 
W. Forman$^{4}$, C. Jones$^{4}$, B. McNamara$^{4}$ \\
\footnotesize
1. Institute of Astronomy, Madingley Road, Cambridge CB3 0HA\\
2. NRAO, P.O. Box O, Socorro, NM, 87801, USA\\
3. Department of Engineering Physics, University of Wollongong, Wollongong NSW 2522, Australia\\
4. Harvard-Smithsonian Center for Astrophysics, 60 Garden Street, Cambridge MA 02138, USA
}}
\begin{document}
\maketitle
\begin{abstract}
We examine the properties of the X-ray gas in the central regions of 
the distant ($z=0.46$), X-ray luminous cluster of galaxies 
surrounding the powerful radio source 3C295, using observations made with 
the Chandra Observatory. Between radii of $50-500$ kpc, the cluster gas 
is approximately isothermal with an emission-weighted 
temperature, $kT\sim 5$ keV. Within the central 50 kpc radius 
this value drops to $kT \sim 3.7$ keV. The spectral and imaging Chandra 
data indicate the presence of a cooling flow within the central 
50 kpc radius of the cluster, with a mass 
deposition rate of approximately 280 \Msunpyr. 
We estimate an age for the cooling flow 
of $1-2$ Gyr, which is approximately one thousand times older 
than the central radio source. We find no evidence in the X-ray spectra 
or images for significant heating of the X-ray gas by the radio source.
We report the detection of an edge-like absorption feature in the 
spectrum for the central 50 kpc region, which may be due to oxygen-enriched 
dust grains. The implied mass in metals seen in absorption could have been 
accumulated by the cooling flow over its lifetime. Combining the results on 
the X-ray gas density profile with radio measurements of the Faraday 
rotation measure in 3C295, we estimate the magnetic field strength 
in the region of the cluster core to be $B \sim 12 \mu$G.  
\end{abstract}

\begin{keywords}
galaxies: active -- 
galaxies: clusters: individual: 3C295  -- cooling flows -- 
intergalactic medium -- radio continuum: galaxies -- X-rays: galaxies
\end{keywords}

\section{Introduction}

X-ray observations of clusters of galaxies show that in the central
regions of many clusters the cooling time of the intracluster gas
is significantly less than a Hubble time (\eg White, Jones \& Forman
1997; Peres \etal 1998). This cooling is thought to lead to a slow net 
inflow of material towards the cluster centre; a process known as a 
cooling flow (see Fabian 1994 for a review). X-ray imaging data show 
that the gas typically `cools out' and is deposited throughout the central 
few tens to hundreds of kpc in clusters, with ${\dot M}(r) \approxpropto r$, 
where ${\dot M}(r)$ is the integrated mass deposition rate within radius 
$r$ (\eg Thomas, Fabian \& Nulsen 1987). Spatially resolved X-ray spectroscopy 
of nearby clusters with ROSAT and ASCA has confirmed the presence of 
distributed, relatively cool gas in cooling flows, with 
spatial distributions and luminosities in good agreement with the 
predictions from the imaging data and  cooling flow models (\eg 
Allen \& Fabian 1997; Allen \etal 2000). 

In this paper we present Chandra (Weisskopf \etal 2000) X-ray observations of the 
central region of the distant ($z=0.4605$), optically rich 
(Dressler \& Gunn 1992), X-ray luminous  (Henry \& Henriksen 1986) cluster 
of galaxies surrounding the powerful radio source 3C295. Previous 
X-ray imaging studies of the cluster with the Einstein Observatory and 
ROSAT have suggested that the cluster contains a strong cooling flow 
in the region surrounding its central radio source 
(Henry \& Henriksen 1986; White \etal 1997; Neumann 1999). 
Here we present the first spatially resolved X-ray spectroscopy of the 
cluster which support this hypothesis. 
We estimate the age of the cooling flow using the Chandra data and show that it is 
approximately one thousand times older than the central radio source. 

Harris \etal (2000) have reported the detection of X-ray emission 
from the radio hot-spots in 3C295, using the same Chandra observations 
discussed here, which they associate with synchrotron self-Compton emission. 
We here extend the combined radio/X-ray analysis and estimate the 
magnetic field strength in the cluster core.

The cosmological parameters $H_0$=50 \kmpspMpc, $\Omega = 1$ and 
$\Lambda = 0$ are assumed throughout. At the redshift of 3C295 
($z=0.4605$), an angular scale of 1 arcsec corresponds to a physical size 
of 6.87 kpc.

\section{Chandra observations and data reduction}

The Chandra observations of 3C295 were made using the Advanced CCD
Imaging Spectrometer (ACIS) on 1999 August 30, during the
performance verification and calibration phase of the mission. The
target was observed in the back-illuminated CCD detectors, 
close to the nominal aim point for the S3 detector. (The source
was positioned close to the node-0/node-1 boundary of the chip). The 
focal plane temperature at the time of the observations was -100C.

We have used the level-2 events file provided by the standard Chandra pipeline
processing and the $CIAO$ software available from the 
Chandra X-ray Center (CXC; {\it http://asc.harvard.edu/ciao/} for 
our analysis. Further manual screening based on the 
instrument light curve was also carried out to remove periods 
of anomalously high background at the beginning and end of the observation. Only
those X-ray events with grade classifications of 0,2,3,4 and 6 
were included in our final cleaned data set, which had an exposure time of 17.0ks.

\section{Basic imaging analysis}

\subsection{X-ray and radio morphology}

\begin{figure*}
\vspace{2cm}
\hbox{
\hspace{1.0cm}\psfig{figure=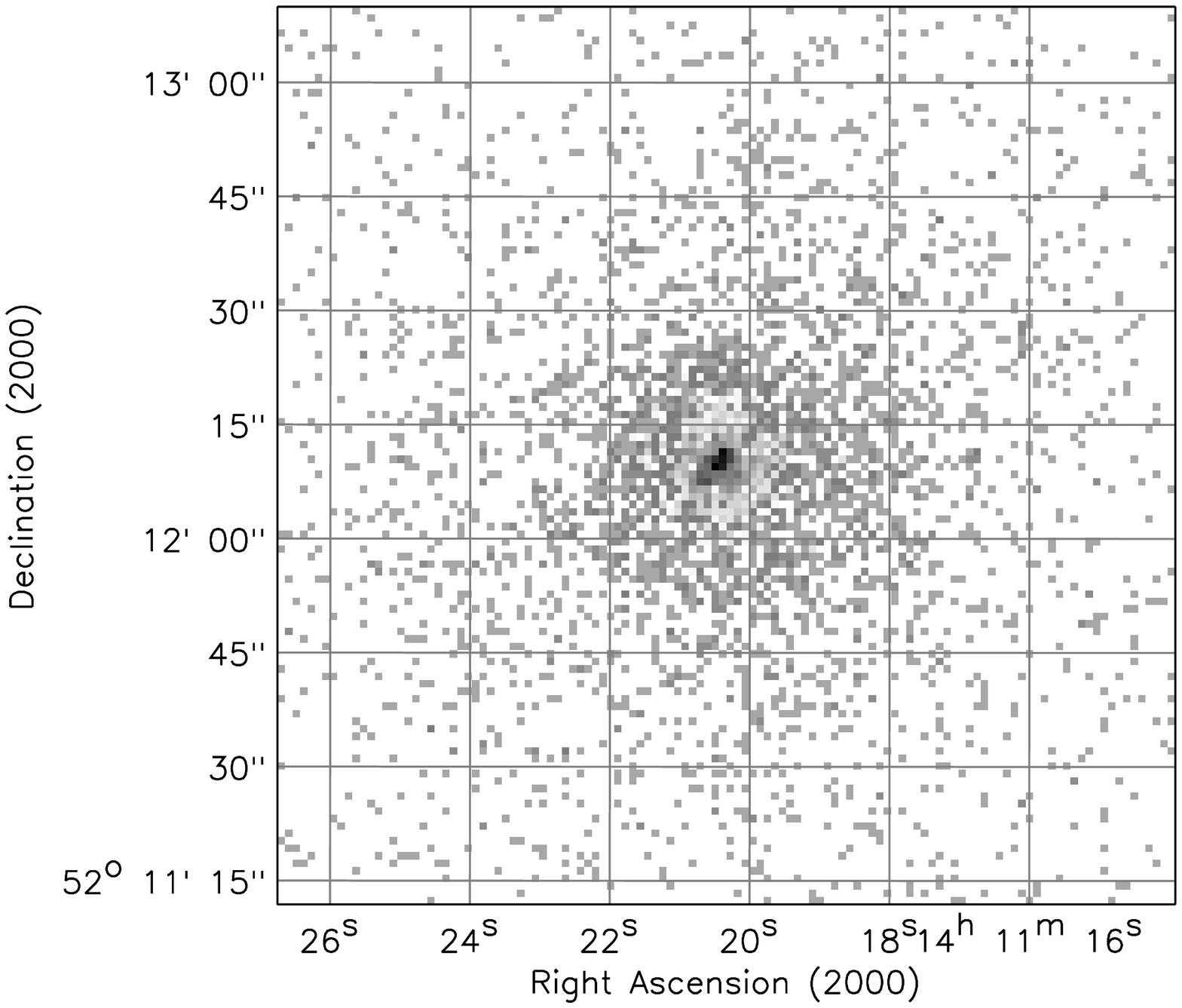,width=0.45 \textwidth,angle=0}
\hspace{0.8cm}\psfig{figure=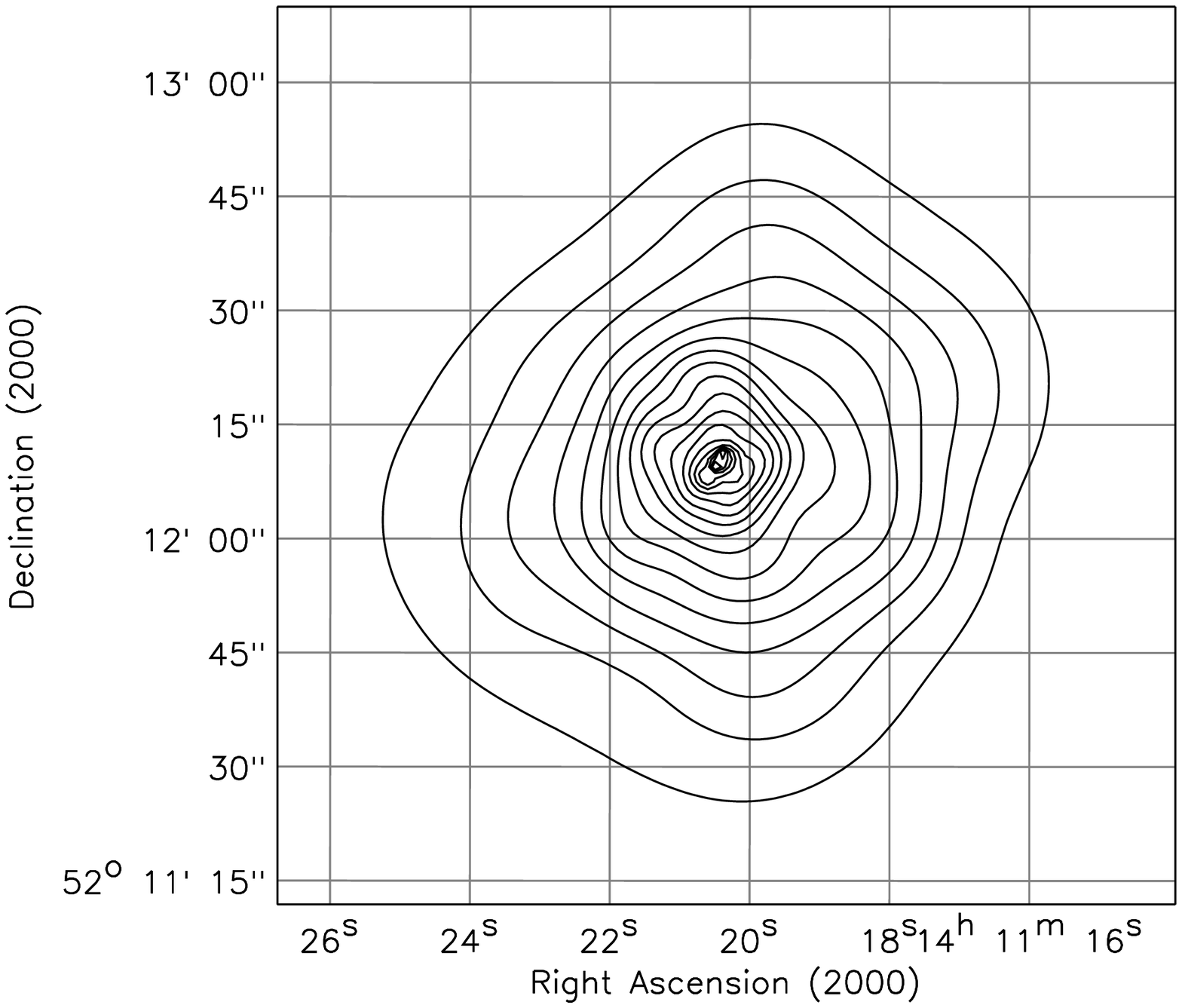,width=0.45 \textwidth,angle=0}
}
\caption{(Left panel) The raw $0.3-7.0$ keV Chandra image of 3C295. 
The pixel size is 2 detector pixels (0.9838 arcsec). (Right panel) Contour 
plot of the same region, adaptively smoothed using the code of Ebeling, 
White \& Rangarajan (2001), with a 
threshold value of $3.5 \sigma$. 
The contours have equal logarithmic spacing. }\label{fig:im1} 
\end{figure*}

The raw $0.3-7.0$ keV image for the central $2 \times 2$ 
arcmin$^2$ ($0.825 \times 0.825$ Mpc$^2$) region of the 3C295 
cluster is shown in Fig. \ref{fig:im1}(a). The pixel size is 
$0.984  \times 0.984$ arcsec$^2$, corresponding to 
$2 \times 2$ raw detector pixels. Fig. \ref{fig:im1}(b) shows 
an adaptively smoothed contour plot of the same data, using the 
smoothing algorithm of Ebeling, White \& Rangarajan (2001). 

\begin{figure*}
\hbox{
\hspace{3cm}\psfig{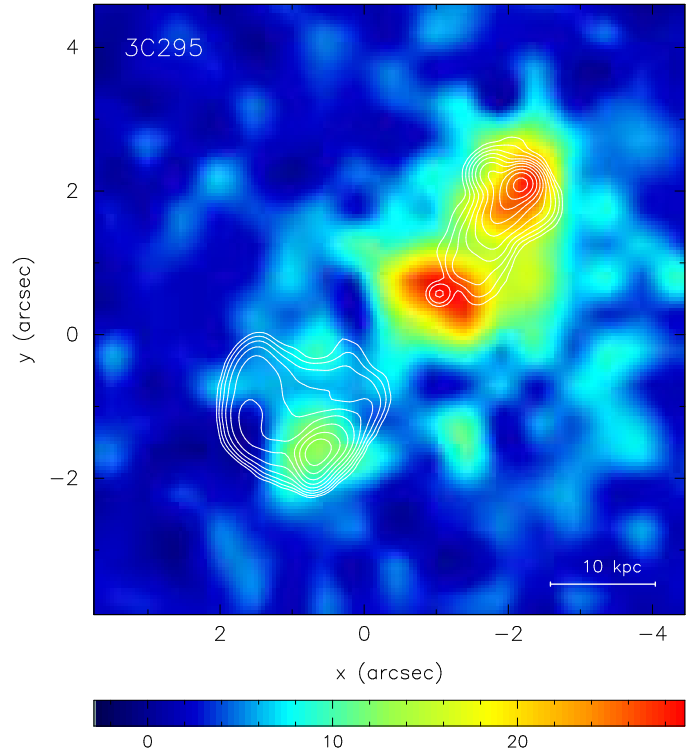}
}
\caption{The 8.4 GHz radio image of 3C295 (contours) overlaid on the 
Chandra image at maximum spatial resolution ($0.492  \times 0.492$ arcsec$^2$ pixels).
The central radio source and hot-spots are clearly detected in X-rays 
(see also Harris \etal 2000).}
\label{fig:radio_image}
\end{figure*}

Extensive multi-frequency radio observations of 3C295 have been
carried out by Perley \& Taylor (1991) and Taylor \& Perley (1992).
These observations reveal a compact (40 kpc diameter) double lobed 
radio galaxy with prominent steep spectrum hot spots on both sides 
of a weak flat spectrum core. Fig. \ref{fig:radio_image} shows an 
8.4 GHz radio image of the source (contours) overlaid on the $0.3-7.0$ keV 
Chandra image (in this case shown at maximum spatial resolution with 
$0.492  \times 0.492$ arcsec$^2$ pixels). The position 
for the  8.4 GHz radio core is 14h11m20.53$\pm0.01$ +52d12m09.69$\pm0.05$s (J2000.) 
The Chandra image has been shifted by 0.015 arcsec east and 
0.377 arcsec north (these shifts are within the astrometry errors) which 
provides excellent agreement between the positions of the radio 
and X-ray core and hot-spots. The central radio source and both radio 
lobes are clearly detected in X-rays, which Harris \etal (2000) 
attribute to synchrotron self-Compton emission. 

We have searched for the presence of additional patchy substructure 
in the Chandra images by comparing the data with simple
elliptically-symmetric models. The data do not exhibit 
substructure on scales $>1$ arcsec that is significant at the 
$>3 \sigma$ level, other than that associated with the central radio source.
The radio data for the cluster are discussed in more detail in 
Section 6.

\subsection{The surface brightness profile}

\begin{figure}
\vspace{2cm}
\hbox{
\hspace{0.0cm}\psfig{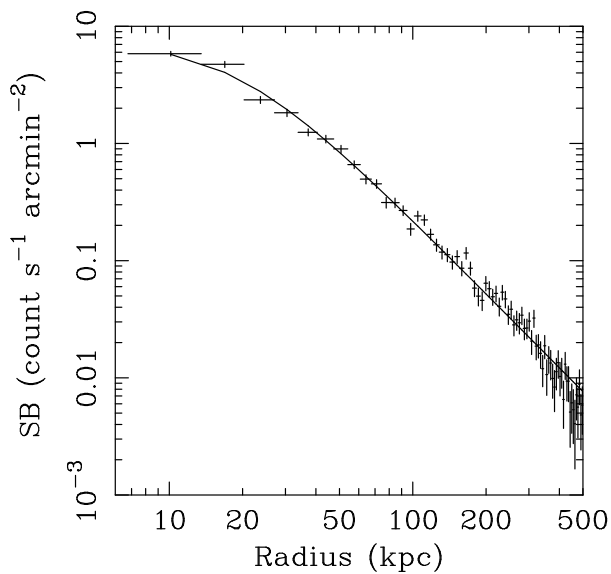}
}
\vspace{1cm}
\caption{The background-subtracted, azimuthally-averaged radial surface 
brightness profile for the central 500 kpc region of the 
3C295 cluster in the $0.3-7.0$ keV band. The binsize is 0.984 arcsec 
(6.76 kpc). All point sources, including the central AGN, and the 
regions of enhanced emission associated with the radio lobes have been excluded.}
\label{fig:surbri}
\end{figure}

X-ray emission from the 3C295 cluster is detected out 
to a radius $r \sim 1$ Mpc (2.4 arcmin) in the Chandra 
data. Beyond $r \sim 500$ kpc (1.2 arcmin), however, 
background counts dominate the detected flux. Fig. \ref{fig:surbri} 
shows the azimuthally-averaged, $0.3-7.0$ keV surface brightness profile 
for the central 500 kpc radius. The bin-size is 2 
detector pixels (0.984 arcsec). The data have been background subtracted 
using a rectangular region of size $5 \times 1$ arcmin$^2$, located 
$\sim 5.7$ arcmin (2.35 Mpc) from the cluster centre. All point 
sources, including the central AGN and regions of enhanced emission 
associated with the radio lobes have been excluded. 

Within a radius $r \sim500$ kpc (excluding the innermost bin associated 
with the central AGN), the X-ray surface brightness profile can be 
approximated ($\chi^2 = 101$ for 70 degrees of freedom) by a 
an isothermal $\beta$-model (\eg Jones \& Forman 1984) of the 
form $S(r) = S(0)\left [ {(1+{r/r_{\rm c})}^2} \right ]^{1/2-3\beta}$, 
with a core radius $r_{\rm c} = 18.8\pm1.0$ kpc and a slope parameter 
$\beta =0.517\pm0.005$ ($1 \sigma$ errors; $\Delta \chi^2=1.0$). 
The best-fitting $\beta$-model is shown overlaid on the surface brightness
data in Fig. \ref{fig:surbri}.

\subsection{X-ray `colour' profile}

\begin{figure}
\vspace{2cm}
\hbox{
\hspace{0.0cm}\psfig{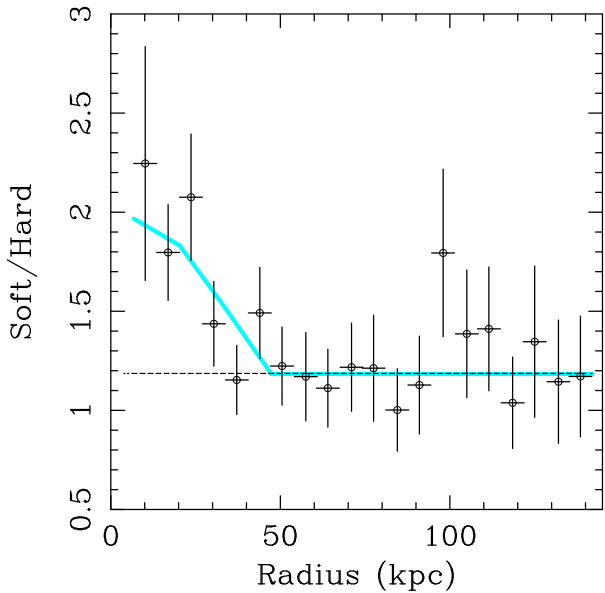}}
\vspace{1cm}
\caption{The X-ray colour, defined by the ratio of the counts in the 
$0.5-1.2$ and $1.3-7.0$ keV bands (soft/hard), as a function of radius. 
At large radii the ratio is approximately constant and 
consistent with a temperature $kT =5.0^{+1.3}_{-0.9}$ keV. Within a `break' 
radius of $53^{+22}_{-19}$ kpc, however, the colour ratio rises sharply indicating
the presence of cooler gas. 
Comparison with the cooling time curve shown in Fig. \ref{fig:deproj}(b) 
suggests an age for the cooling flow of $1-2$ Gyr. 
The grey curve shows the predicted colour profile for a cooling 
flow of age $1.5$ Gyr, assuming that the gas cools from an upper temperature 
$kT \sim 5$ keV
within the central 50 kpc. A metallicity of 0.5 solar and a Galactic column 
density of $1.3 \times 10^{20}$\apc~are assumed.}\label{fig:colour}
\end{figure}

In order to investigate radial variations in the temperature 
of the X-ray gas at high spatial resolution, 
we have constructed an X-ray colour profile for the central regions 
of the cluster. Two separate images were created 
using counts in Pulse Invariant (PI) channels $36-83$ and 
$84-480$, which cover the observed energy ranges 
$0.5-1.2$ and $1.2-7.0$ keV (or $0.73-1.75$ and $1.75-10.2$ 
keV in the rest frame of the source). The soft and hard X-ray images 
were background subtracted and all significant point sources, including 
the central AGN and the brightest regions of X-ray emission 
associated with the radio lobes, 
were masked out and excluded from the analysis.
Azimuthally-averaged surface brightness profiles 
for the cluster were then constructed
in each energy band, centred on the overall peak of the 
X-ray emission (Section 3.1). The X-ray `colour' profile, 
formed from the ratio of the surface brightness profiles in
the soft and hard bands, is shown in Fig. \ref{fig:colour}. 

\begin{figure*}
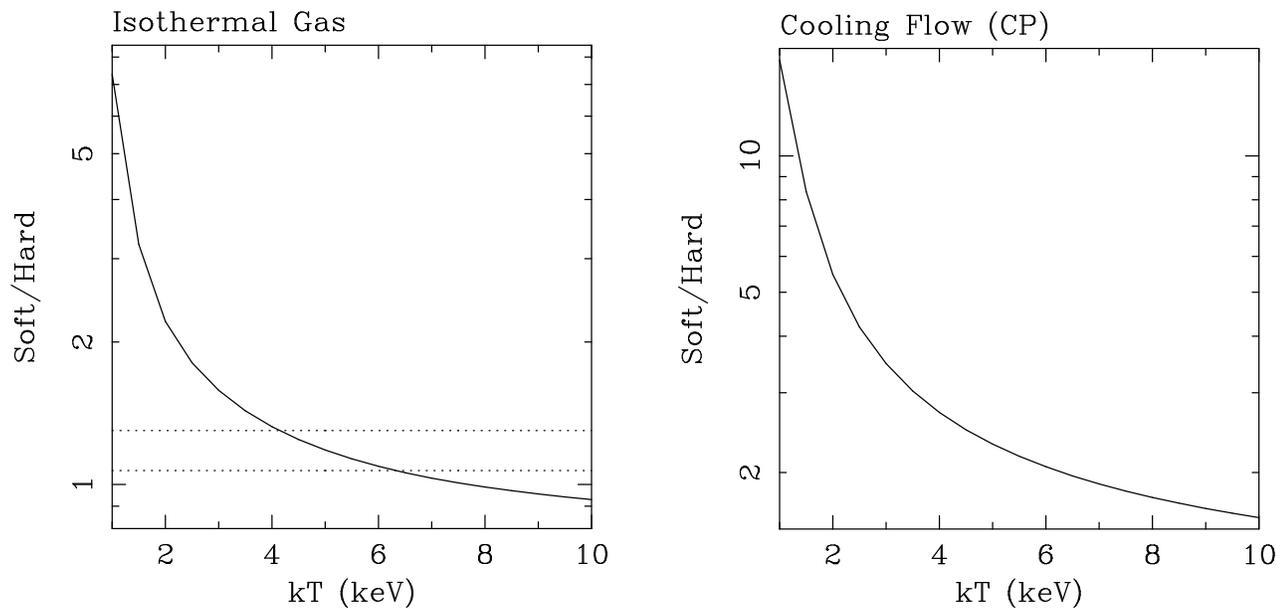

\vspace{2cm}
\hbox{
\hspace{0.0cm}\psfig{figure=theory_iso_newrat_tidy.ps,width=0.45 \textwidth,angle=270}
\hspace{0.8cm}\psfig{figure=theory_cflow_newrat_tidy.ps,width=0.45 \textwidth,angle=270}
}
\vspace{1cm}
\caption{Theoretical expectations for the X-ray colour ratios of
isothermal gas (left panel) and a constant pressure cooling flow
(right panel) as a function of temperature. The dotted lines in the
left panel show the range of values consistent with the observed
colour ratio between radii of $50-140$ kpc in 3C295. In the right panel, the
temperature is the upper temperature from which the gas cools. A
metallicity of 0.5 solar, an absorbing column density of $1.3 \times
10^{20}$\apc~and a redshift of 0.4605, are assumed.}\label{fig:theory} 
\end{figure*}

Fig. \ref{fig:theory} shows the theoretical expectations 
for the X-ray colour for both isothermal gas and a constant pressure cooling
flow, as a function of temperature. (A metallicity of 0.5 solar and a
Galactic column density of $1.3\times 10^{20}$\apc~are assumed.) For
the cooling flow predictions, the temperature refers to the upper 
temperature from which the gas cools). We see that the observed X-ray 
colour ratio of $1.19\pm0.12$ in the 
outer regions of 3C295 (determined from a $\chi^2$ fit with a
constant model in the $50-140$ kpc range) implies a gas temperature of
$5.0^{+1.3}_{-0.9}$ keV (in excellent agreement with the results from 
the spectral analysis discussed in Section 4.3.1). Within a `break' radius 
of $53^{+22}_{-19}$ kpc,  however, the colour ratio rises sharply, 
indicating the presence of cooler gas (the $1\sigma$ errors on the break 
radius are determined from a $\chi^2$ fit to the data in Fig. 
\ref{fig:colour}~with a broken power-law model). 
We note that the observed X-ray colour ratio is relatively insensitive to variations in 
metallicity within the cluster, ranging  from 1.19 to 1.17 for 
isothermal 5 keV gas as the metallicity is varied from $0.0-1.5$ solar.

\section{Spatially-resolved spectroscopy}

\subsection{The regions studied}

For our spectral analysis, we have divided the cluster into three 
annuli, covering the radial ranges $0-50$ kpc (within which the 
X-ray colour profile shown in Fig. \ref{fig:colour}~
indicates the presence of cooler gas), $50-250$ kpc 
and $250-500$ kpc. Spectra were extracted from each 
region in 2048 Pulse Height Analyser
(PHA) channels, which were re-grouped to contain a minimum of 20 counts
per bin, thereby allowing $\chi^2$ statistics to be used. (For the
outermost annulus, a grouping of 80 counts per PHA bin was used due to
the larger background contribution.) A background spectrum was extracted from a 
rectangular region, approximately $480 \times 140$ arcsec$^2$ in size, located
well away from the cluster in nodes 2 and 3 of the S3 chip.\footnote{In practice, a
range of different 
background regions were examined, covering various source-free areas of the S3 
chip. The results presented in this paper are not sensitive to the precise 
choice of background region.} 
All significant point sources, including the central AGN, 
were masked out and excluded from the analysis. In addition, for the 
central 50kpc region, the X-ray emission associated with the radio lobes 
(Harris \etal 2000) was excluded using two $2\times2$ arcsec$^2$ square masks.
(The X-ray emission from the central AGN and radio lobes are  
discussed in Section 4.5.)

Separate photon-weighted response matrices and 
effective area files were constructed for each annular region 
studied using the calibration and response files appropriate 
for the focal plane temperature, available from the 
CXC.\footnote{For each $32 \times 32$ pixel$^2$ sub-region of the S3 chip, 
a spectral response (.rmf) and an auxiliary response (.arf) matrix were 
created using the CIAO tools $mkrmf$ and $mkarf$, respectively. For each 
of the three annular regions studied, the number of source counts 
contributed from each $32 \times 32$ pixel$^2$ sub-region was determined. 
The individual .rmf and .arf files were then combined (using the FTOOLS 
programs $addrmf$ and $addarf$) to form a counts-weighted spectral 
response and auxiliary response matrix appropriate for each annulus.} 
Two separate energy ranges were analysed:  
a conservative $0.5-7.0$ keV range, over which the calibration of
the back-illuminated CCDs is currently best understood, and  
a more extended $0.3-7.0$ keV energy range, 
which provides extra information on cool emission components and 
intrinsic absorption in the cluster.

\subsection{The spectral models}

The analysis of the spectral data has been 
carried out using the XSPEC software package 
(version 10.0; Arnaud 1996). The spectra 
were modeled using the plasma emission code of Kaastra \& Mewe
(1993; incorporating the Fe L calculations of Liedhal, Osterheld \&
Goldstein 1995) and the photoelectric absorption models of 
Balucinska-Church \& McCammon (1992).
We first examined each annular 
spectrum using a simple, single-temperature model with the absorbing column
density fixed at the nominal Galactic value ($N_{\rm H} = 1.33 \times
10^{20}$\apc; Dickey \& Lockman 1990). This model is hereafter
referred to as model A. The free parameters in model A were the 
temperature ($kT$) and metallicity ($Z$) of the plasma 
(measured relative to the solar photospheric values of Anders 
\& Grevesse 1989, with the various elements assumed to be present 
in their solar ratios) and the emission measure ($K$). We also examined 
a second single-temperature model (model B) which was 
identical to model A but with the absorbing column 
density ($N_{\rm H}$; assumed to act at zero redshift) 
also included as a free parameter in the fits.  

Motivated by the results on the X-ray colour profile shown in 
Fig. \ref{fig:colour} 
and the results from the deprojection analysis discussed in Section 5.3, we
have also examined the spectrum for the central 50 kpc region with a
series of more sophisticated, multiphase emission models in which 
the properties of any cooling flow present can be explicitly accounted for.
In the first such model, model C1, the cooling gas was assumed to cool 
at constant pressure from the ambient cluster temperature, following the 
prescription of Johnstone \etal (1992). In the second case, model C2, the 
cooling flow was modeled as an `isothermal' cooling flow, following Nulsen 
(1998)\footnote{In the `isothermal' cooling flow model it is assumed that the 
distribution of temperature and density inhomogeneities with radius is 
self-similar and that the mean gas temperature remains constant. We assume a 
value for $\eta =1$, where the integrated mass deposition rate within 
radius r, ${\dot M} \propto r^{\eta}$}. In both cooling-flow models, the 
normalization of the cooling-flow component was parameterized in terms
of a mass deposition rate, ${\dot M}$, which was a free parameter in the 
fits. Finally, we also examined a more general emission model,
model D, in which the cooling gas was modelled by a second, cooler
isothermal emission component, with the temperature and 
normalization of this component included as free fit parameters. 
Model D provides a more flexible parameterization, with an 
additional degree of freedom over models C1 and C2, and 
invariably provides a good match to the more specific 
cooling-flow models at the spectral resolution and 
signal-to-noise ratios typical of ACIS observations of distant clusters.
However, we found that the ambient cluster temperature was 
not well constrained with model D, and thus we do not quote explicit 
results for this model here.

With each of the cooling-flow emission models, we have also examined the 
effects of including extra absorption, 
using a variety of different absorption models. In the first
case (absorption model i), the only absorption included
was that due to cold gas in our Galaxy, with the equivalent 
column density fixed to the nominal Galactic value 
(Dickey \& Lockman 1990; For a single temperature emission model, 
this would be identical to spectral model A). In the second
model (model ii), the absorption was again assumed to be 
due to Galactic (zero redshift) cold gas, 
but with the column density, $N_{\rm H}$, included as a free 
parameter in the fits. (For a single temperature emission model, 
this is equivalent to spectral model B.) In 
the third case (absorption model iii), 
an intrinsic absorption component with column density,
 $\Delta N_{\rm H}$, due 
to cold gas at the redshift of the cluster was introduced. 
The absorber was assumed to lie in a uniform screen in front of the
cooling flow, with the column density included as a free fit parameter.  
In the fourth case (model iv), the intrinsic absorption was assumed to
cover only a fraction, $f$,  of the emission from the cooling flow.
The fifth and final absorption model (model v) was similar to model (iii) 
but with the gaseous absorber replaced by an intrinsic absorption 
edge, with the edge depth, $\tau$, and energy, $E_{\rm edge}$, 
free parameters in the fits. This general absorption model may be used to 
approximate the effects of a dusty and/or ionized absorber.

In those cases where the absorption has been quantified in terms of  
an equivalent hydrogen column density, solar metallicity in the 
absorbing gas is assumed.
We note that in absorption
models (iii--v), the absorption acting
on the ambient cluster emission was fixed at the nominal Galactic
value. However, allowing the Galactic absorption to vary from this value did
not significantly improve the fit (at $\geq 95$ per cent confidence) in any 
case.

\subsection{Results from the spectral analysis}

The best fit parameter values and
90 per cent ($\Delta \chi^2 = 2.71$) confidence limits determined from
the spectral analysis are summarized
in Tables $1-3$. For the multiphase analysis of the central 50 kpc
region over the restricted $0.5-7.0$ keV band, only the results for 
absorption models (i) and (ii) are listed, since the other  
models were poorly constrained. A complete summary of results 
from the multiphase analysis using the full $0.3-7.0$ keV energy band is
provided in Table 3. 

\subsubsection{Single temperature emission models}

Figs. $6-8$ show the Chandra spectra and best-fitting
single-temperature models (using model B) for the three annular regions. 
In all cases, the fits are acceptable in terms of the $\chi^2$ values 
obtained (although for the central 50kpc region, the fits were significantly 
improved using the multiphase emission plus absorption edge models discussed 
in Section 4.3.2 below). The results determined in the $0.5-7.0$ and 
$0.3-7.0$ keV energy ranges show good agreement, arguing against serious 
calibration problems at low energies in the data. The results for the 
$50-250$ and $250-500$ kpc annuli both indicate an ambient cluster
temperature of $\sim 5$ keV. A joint fit to the data for the two outer annuli 
with spectral model B over the $0.3-7.0$ keV band gives a mean, ambient 
cluster temperature, $kT= 5.3^{+1.2}_{-0.7}$ keV.

\begin{figure}
\vspace{2cm}
\hbox{
\hspace{0.0cm}\psfig{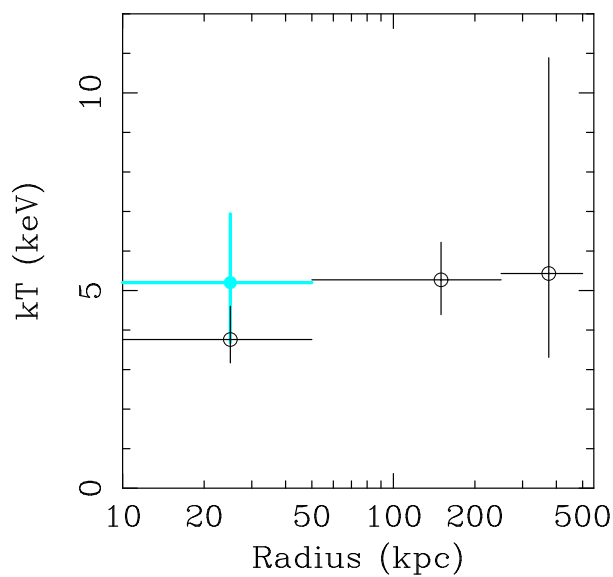}
}
\vspace{1cm}
\caption{The X-ray gas temperature (and 90 per cent confidence limits) 
as a function of radius, determined from the $0.3-7.0$ keV data. Open 
circles show the emission-weighted results determined with the  
single-temperature model (model B). The ambient temperature 
in the central 50 kpc region, determined from the multiphase analysis 
with the effects of the cooling flow and absorption edge accounted for 
(Section 4.3.2), is shown as the solid grey point.}\label{fig:kt}
\vskip 1cm
\end{figure}

Within the central 50 kpc region, the results determined with the single
temperature models (A and B) indicate a sharp drop
in the mean emission-weighted temperature to a value of
$\sim 3.7$ keV. (The temperature profile for the X-ray gas determined
with spectral model B in the $0.3-7.0$ keV band is shown in Fig. 
\ref{fig:kt}.) We find no evidence for intrinsic absorption 
due to cold gas at any radius using the single temperature models and for
all three annuli the  best-fit column densities 
measured with spectral model B are in good agreement with the nominal 
Galactic value of $1.3 \times 10^{20}$\apc. The results indicate a rather 
high metallicity, $Z\sim0.6Z_{\rm \odot}$, across the central 250 
kpc of the cluster. (The metallicity at larger radii is not firmly 
constrained.)

\subsubsection{Multiphase emission models}

\begin{figure}
\vspace{2cm}
\hbox{
\hspace{0cm}\psfig{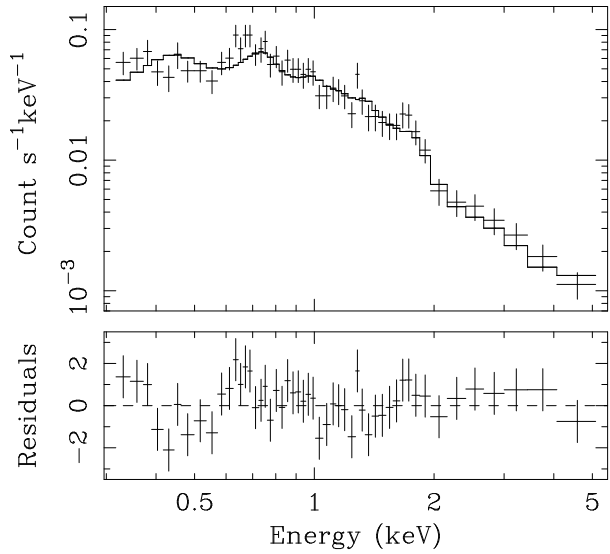}
}
\vspace{1cm}
\caption{(Upper panel) Chandra spectrum for the central 50 kpc
region of 3C295 in the $0.3-7.0$ keV band with the
best-fit single-temperature
plasma model (model B) overlaid.
(Lower Panel) The residuals to the fit in units of $\chi$.} 
\label{fig:spec_core}
\end{figure}

Although the fits to the central 50kpc with spectral models A and B 
are formally acceptable in terms of their $\chi^2$ values, the residuals 
shown in Fig. \ref{fig:spec_core} indicate a systematic problem at low 
energies. The data for the outer annuli do not exhibit such poor residuals 
despite, in the case of the $50-250$ kpc region, having a higher 
signal-to-noise ratio. We are therefore motivated to examine the 
improvements in the fits to the spectrum for the central 50 kpc that may be 
obtained using the more sophisticated, multiphase emission and absorption 
models described in Section 4.2. 

\begin{figure}
\vspace{2cm}
\hbox{
\hspace{0.0cm}\psfig{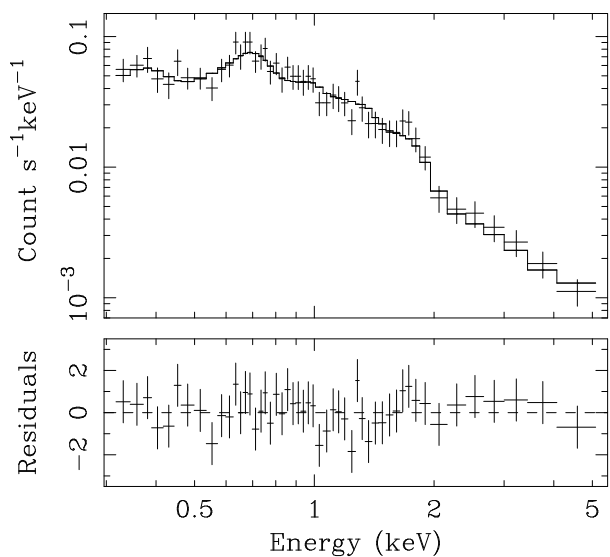}}
\vspace{1cm}
\caption{Chandra spectrum for the central 50 kpc
region of 3C295 in the $0.3-7.0$ keV band with the best-fitting
cooling flow model incorporating an intrinsic absorption edge 
(model C2v) overlaid. Note the improvement to the fit at 
low energies with respect to Fig. \ref{fig:spec_core}.}\label{fig:spec_core_c2}
\end{figure}

The fits with the multiphase cooling flow models incorporating variable 
zero-redshift or intrinsic absorption due to cold gas
(absorption models (ii) and (iii)) provide only marginal improvements in 
$\chi^2$ with respect to model A(i). A significantly better fit  
with the cooling flow emission models ($\Delta \chi^2 \sim 17$) 
is obtained, however, when we also include an absorption edge 
(absorption model v), with an 
intrinsic edge energy, $E_{\rm Edge} \sim 0.6$keV and an optical depth, $\tau \sim 1$.
The best-fit model and data, and the residuals to the fit, are 
shown in  Fig. \ref{fig:spec_core_c2}. The improvement in the fit at low 
energies obtained with cooling flow-plus-edge-absorption model 
may be seen by comparison with Fig. \ref{fig:spec_core}.

We note that a single-temperature emission model incorporating an
absorption edge (model A(v) in Table 3 ) does not provide as good
a fit. Thus, both a multiphase emission model
and an edge-like absorption model are favoured by the data for the 
central 50kpc region. 

We adopt the results determined with spectral
model C2(v) as our preferred values, since this is the most 
physically-motivated model. The results determined with model C2 
show that the
cluster contains a relatively strong cooling flow in its central 50kpc 
region, with a  mass
deposition rate of $281^{+85}_{-95}$ \Msunpyr. The cooling flow
accounts for essentially the entire flux from the central 50 kpc, 
as expected within the context of the `isothermal' cooling flow model, 
given the results on the probable size of the cooling flow in Section 5.3. 
The ambient temperature in the central region of the cluster, 
corrected for the effects of the cooling flow, is 
$5.20^{+1.74}_{-1.52}$ keV (shown as the filled circle in Fig. \ref{fig:kt}),
in good agreement with the value measured at larger radii. This indicates 
that the ambient temperature in the cluster remains approximately
isothermal within the central 500 kpc. The metallicity within the central
50 kpc measured with model C2(v), $Z=0.47^{+0.35}_{-0.26} Z_{\odot}$, 
is in reasonable agreement with the result determined with the simple, 
single temperature models. 

Assuming that the absorption edge is 
due to oxygen, the element with the largest photoelectric absorption cross 
section  ($\sigma \sim 5 \times 10^{-19}$cm$^2$) in this region of the X-ray 
spectrum, the best-fit edge depth of $\tau = 1.14^{+0.97}_{-0.41}$ 
corresponds to an oxygen column density, $\Delta N_{\rm O} = 2.3^{+2.0}_{-0.9} 
\times 10^{18}$\apc~and/or an equivalent hydrogen column density, 
$\Delta N_{\rm H} = 2.7^{+2.3}_{-1.0}  \times 10^{21}$\apc. 
(An oxygen number density relative to hydrogen equal to the solar value 
of $8.51\times 10^{-4}$ is assumed; Anders \& Grevesse 1989). The apparent edge energy of 
$0.57-0.65$ keV is nominally well-matched to the K-absorption edge of 
partially ionized oxygen, with an ionization state in the range  OIII--OV.  However, 
the gain calibration at low energies ($E<0.5$ keV) is 
presently uncertain (Section 4.7 ) and the apparent edge energy is 
probably consistent with any ionization state of oxygen from OI to OV. 
(The photoelectric absorption cross section for oxygen 
does not vary greatly for any ionization state from OI-OV).

\subsection{Spectral deprojection analysis}

The results discussed in Section 4.3 are based on
the analysis of projected spectra. In order to estimate the effects 
of projection, we have also carried out a further analysis of the 
Chandra data using a simple spectral deprojection technique. 
(A similar method is described by Buote 2000a). 

A series of spectra were extracted for the nested set of annuli 
defined by $r_1 < r_2 \ldots < r_n$, which were then used 
to determine the mean, emission-weighted temperatures and 
densities for the set of spherical shells defined by the same set of radii.
The emission from the annulus between $r_i$ and $r_{i+1}$ is made up of
contributions from the spherical shell between $r_i$ and $r_{i+1}$ and
all outer shells. We assume that the emission from each spherical shell is 
isothermal and absorbed by the Galactic column density.
Thus, the spectral model for the annulus between $r_i$ and $r_{i+1}$ is a 
weighted sum of $n-i$ absorbed, isothermal models. The fit to the outermost 
annulus determines the parameters for the outermost spherical shell, which are
held fixed in the spectral models for all remaining annuli.  The
contribution of the outermost shell to each inner shell is determined
by purely geometric factors. The fit to the second annulus inward
determines the parameters for the second spherical shell, and so
forth, working inward. (This parallels the usual image
deprojection procedure.) Fitting of the spectral shells was carried 
out in the $0.5-7.0$ keV band.

We have studied six annuli covering the central 2 arcmin of the cluster.
No attempt was made to correct for residual emission from the X-ray 
gas at larger radii since the steeply rising surface brightness profile of the 
cluster means that only the outer 1 or 2 rings will be significantly 
affected.  (The results for these shells should therefore be viewed 
with caution, although the temperatures in these regions are poorly 
determined in any case). The temperature profile determined from the 
spectral deprojection analysis is shown in Fig. \ref{fig:kt_deproj}. 
The results are in good agreement with those determined from the projected 
spectra (Section 4.3.1). We therefore conclude that projection effects do not 
severely affect the results from the spectral analysis.

\subsection{X-ray emission from the central AGN and radio lobes}

We have extracted X-ray spectra for the central AGN and the brighter 
(northwest) of the two radio lobes. In both cases, a square extraction 
region of size $3 \times 3$ pixel$^{2}$ 
(approximately $1.5 \times 1.5$ arcsec$^2$) was used.
The local background was determined from a circular region of 
radius 2.7 arcsec, centred on the AGN, with the emission from the AGN and 
both radio lobes excluded. 

The background-subtracted count rates for both sources 
are rather low, with a total of 95 and 70 counts 
for the AGN and northwest lobe, respectively. 
Fitting a simple power-law model to the spectra, with the column density fixed to the 
Galactic value of $1.3 \times 10^{20}$\apc, we determine best-fit photon 
indices of $\Gamma = -0.2 \pm 0.4$ for the AGN and $\Gamma = 2.7 \pm 0.8$ for 
the X-ray emission associated with the northwest radio lobe. 
The $0.5-2.0$ and $2.0-10.0$ keV model fluxes are $6.2 \pm 2.4 
\times 10^{-15}$ \ergpcmsqps ($0.5-2.0$ keV) and $2.1 \pm 0.7 \times 
10^{-13}$ \ergpcmsqps ($2.0-10.0$ keV) for the AGN, and $1.2 \pm 0.4 
\times 10^{-14}$ \ergpcmsqps ($0.5-2.0$ keV) and $6^{+9}_{-5} \times 
10^{-15}$ \ergpcmsqps ($2.0-10.0$ keV) for the northwest lobe. 
The very flat slope for the X-ray spectrum associated with the AGN 
suggests that the emission from the nucleus is heavily obscured and that 
the observed flux is primarily due to X-ray photons Compton-scattered 
into our line of sight by cold, optically thick material close to the 
central X-ray source (\eg Lightman \& White 1988; George \& Fabian 1991; 
Matt, Perola \& Piro 1991).

\subsection{Comparison with previous results}

Our results on the properties of the cluster gas, central AGN and northwest 
radio lobe are in good agreement with those reported by Harris \etal 
(2000). These authors measured a mean, emission-weighted temperature of $4.4 \pm 0.6$ keV 
from a fit with spectral model A to the $0.5-7.0$ keV data from within 90 
arcsec of the nucleus. A similar, combined fit to the
data from all three annuli studied in this paper (which together cover
the central 73 arcsec radius) with spectral model A gives 
$kT=4.61^{+0.55}_{-0.45}$ keV ($0.3-7.0$ keV) or 
$kT=4.37^{+0.50}_{-0.39}$ keV using the $0.5-7.0$ keV data. 
(The temperature and metallicity are linked to take the
same values in each annulus.) This mean, emission-weighted 
temperature is lower than the ambient cluster temperature and 
reflects the presence of cooler gas in the central region of the cluster.

The best-fit ambient cluster temperature of $kT=5.3^{+1.2}_{-0.7}$ keV, determined
from a combined analysis of the Chandra data for the $50-500$ kpc region, 
is slightly cooler than the value 
of $7.1^{+2.1}_{-1.3}$ keV reported by Mushotzky \& Scharf 
(1997), based on ASCA observations. Our own analysis of the 
same ASCA Solid-State Imaging Spectrometer (SIS0) data for the 
central 3.0 arcmin (radius) region of the cluster, using updated 
calibration files and spectral model B, gives a value 
$kT=5.0^{+2.4}_{-1.5}$ keV, with a metallicity of
$0.36^{+0.39}_{-0.32} Z_{\odot}$ and an absorbing column density $N_{\rm H} =
7^{+9}_{-7} \times 10^{20}$ \apc, in good
agreement with the Chandra results. We note that the application of 
standard screening procedures to the ASCA observation of 3C295 excludes 
the bulk of the Gas Scintillation Imaging Spectrometer (GIS) data. 
(We have also excluded the SIS1 data from our analysis, since the 
cluster centre lies close  to the edge of that detector.)

The absorption-corrected bolometric luminosity of the 3C295 cluster 
($L_{Bol} = 2.3\pm0.3 \times 10^{45}$\ergps) determined from
the ASCA data is in good agreement with 
the value of $2.6^{+0.4}_{-0.2} \times 10^{45}$\ergps~reported by
Neumann (1999), based on  ROSAT observations.
Applying the cooling-flow corrected $kT-L_{\rm Bol}$ relation for
cooling-flow clusters determined by Allen \& Fabian (1998), we find
that such a luminosity typically corresponds to a temperature 
of $5.5-6.5$ keV, in good agreement with the  Chandra results.

\subsection{Calibration uncertainties and the apparent absorption edge}

The data for the central 50 kpc region of the cluster statistically
require the introduction of an absorption edge at high formal 
significance. The application of an F-test indicates that the 
improvements obtained with spectral models C1(v) and C2(v), with respect to 
models C1(i) and C2(i), of $\Delta \chi^2 \sim 17$ for the addition of two 
extra fit parameters, are significant at $\gg 99$ per cent confidence. 
However, especially at this early stage in the Chandra mission, it is 
important to consider how calibration uncertainties may also affect 
the results.

The presence of an absorption edge in the data for the central 
50 kpc region is supported by the absence of such a 
feature in the data for the outer annuli.  
Although the introduction of an absorption edge at an energy of 0.63 keV 
with an optical depth $\tau<0.6$ is permitted (though not required) by the 
data for the $50-250$ kpc region, 
the introduction of an edge with an optical depth $\tau = 1.14$ (the best 
fit value for the central 50 kpc) at this same energy leads to a significant 
increase in $\chi^2$ ($\Delta \chi^2=24.1$) with respect 
to the result obtained for no edge. The presence of an edge with the 
best-fit parameter values determined for the central 50 kpc region can 
therefore be ruled out in the $50-250$ kpc data at 
the $4-5\sigma$ level. The data for the 
outer, $250-500$ kpc region are less constraining and 
permit the inclusion of an absorption edge at 0.63 keV with an optical 
depth $\tau < 1$, although the introduction of such an edge provides 
no improvement in $\chi^2$. Thus, the 
absorption signature appears to be concentrated within the central 
50 kpc region of the cluster.

\begin{figure}
\vspace{2cm}
\hbox{
\hspace{0.0cm}\psfig{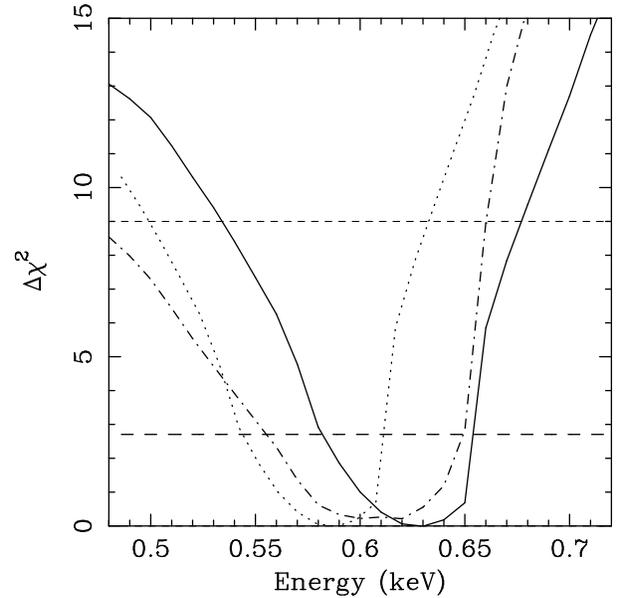}}
\vspace{1cm}
\caption{The increase in $\chi^2$ from the best-fit value as the edge energy 
is varied in spectral model C2(v). The solid curve shows the values determined 
from the PHA analysis with the nominal gain settings. The dotted curve 
shows the same results when  
a gain correction of 7 per cent (the maximum shift consistent with the 
observed energy of the OI\,K fluorescent emission line) is applied. The 
dot-dashed curve shows the results determined from the PI analysis 
(Section 4.7). The horizontal dashed lines mark the 90 per cent (lower curve) 
and $3\sigma$ confidence levels. }\label{fig:edge}
\end{figure}

The intrinsic edge energy of $0.63^{+0.03}_{-0.05}$ keV 
(apparent energy $0.43^{+0.02}_{-0.03}$ keV) determined from 
the analysis of the PHA data with spectral models C1(v) and C2(v) 
is inconsistent with the 
neutral K-edge of oxygen, which should 
occur at an intrinsic energy of $0.53-0.55$ keV (see discussion in 
Arnaud \& Mushotzky 1998).
The solid curve in Fig. \ref{fig:edge} shows 
the relative increase in $\chi^2$ with
respect to the best fit value determined with spectral model C2(v) 
as the intrinsic edge energy is varied between values of 0.5 and 0.7 keV. 
It is crucial to note, however, that the gain 
calibration for the S3 detector is currently only estimated 
to be accurate to within $\sim 5$ per cent in the range $0.3-0.5$ keV; 
N. Schulz, private communication.) To check the low-energy gain calibration, we 
have searched for fluorescent OI\,K emission in the background 
spectrum used in our analysis, which should appear at an energy of 
$0.525$ keV. We indeed detect such a line at an apparent energy of 
$0.552^{+0.014}_{-0.017}$ keV (90 per cent confidence limits) 
which suggests the presence of a gain offset of 
$5.1^{+1.9}_{-3.2}$ per cent at this energy. Applying such a gain 
correction to our results, we recover a true, 
intrinsic edge energy of $0.60^{+0.05}_{-0.06}$ keV, which 
is consistent with the expected edge energy due to neutral oxygen, or 
indeed any ionization state of oxygen up to OV. (The maximally 
gain-corrected results are shown as the dashed curve in Fig. 
\ref{fig:edge}.) 

We have also repeated our analysis of the properties of the 
absorption edge in the central 50 kpc region using a spectrum 
extracted in 1024 Pulse Invariant (PI) detector channels (using a gain file 
appropriate for the observation date). The PI data were grouped to contain a 
minimum of 20 counts per analysis bin, in an identical manner to the PHA 
analysis. The best-fit parameters determined from the PI analysis 
($kT=4.54^{+2.07}_{-1.18}$ keV, $Z=0.47^{+0.36}_{-0.27}$ $Z_{\rm odot}$, 
${\dot M} = 266^{+95}_{-106}$ \Msunpyr, $E_{\rm edge} = 0.60^{+0.05}_{-0.05}$ 
keV and $\tau = 1.59^{+1.99}_{-0.69}$, giving $\chi^2=33.2$ for 42 degrees of 
freedom using spectral model C2(v)) are in good overall agreement with
those determined from the PHA data, although 
the best-fit edge energy in the PI case is consistent with the expected
value for neutral oxygen. (The constraints on the edge energy determined from 
the PI data are shown as the dot-dashed curve in Fig. \ref{fig:edge}.) Taking 
the PI and PHA analyses together, it is apparent that the results on the edge 
energy should be viewed with caution at present, but are probably consistent 
with any ionization state of oxygen from OI-OV. 

Significant gain calibration errors appear 
to be limited to low energies in the S3 detector.
Analysing a single PHA spectrum for the entire central  
500 kpc region, excluding the emission associated with the central 
AGN, we detect a clear Fe-K line complex at an energy of $\sim 4.5$ keV, 
from which we determine a redshift $z=0.461^{+0.019}_{-0.017}$ 
(90 per cent confidence), in excellent agreement with the 
optically-determined result. 

We note that the results on the absorption edge in the central 50 kpc region 
are insensitive to the choice of background region used in the analysis and 
to spatial variations in the spectral response across the 
field of the S3 detector. The fact that the data for the $50-250$ kpc 
annulus exclude the presence of an edge with the best-fit 
parameters determined for the central region, also argues against the 
absorption edge being due to uncertainties in the quantum efficiency 
of the detector at low energies.

Finally, we note one further possibility which may affect the observed 
edge energy, which is resonant scattering of the central AGN continuum by 
OVII gas in the cooling flow. In principle, this could produce an emission 
line at an energy of 0.57 keV with a flux of a few $10^{-15}$ \ergpcmsqps, 
and lead to an apparent shift in the edge energy from 0.54 keV to higher 
values. 

In summary, the presence of a photoelectric absorption edge is statistically 
required by the Chandra data for the central 50 kpc region of the 3C295 
cluster. The edge is not easily explained as an artefact due to calibration 
or modeling errors, although the results on the edge energy should be 
viewed with caution at present. Similar results on the edge energy 
and depth are determined whether we carry out the analysis in PHA or 
PI bins. We note that similar systematic 
residuals at low energies have not, as yet,  been seen by us 
in Chandra data sets for other cluster cooling flows.

\section{Image Deprojection analysis}

We have carried out a deprojection analysis of the Chandra X-ray 
images using an extensively updated version of the deprojection code 
of Fabian \etal (1981; see White, Jones \& Forman 1997 for 
details). For this analysis, we first rebinned the observed 
surface brightness profile (Fig. \ref{fig:surbri}) into 2 arcsec bins. 
For the innermost bin, we have corrected for the excluded cluster 
emission in the region of the central AGN by extrapolating an analytic 
King law fit to the surface brightness profile in the $1-10$ arcsec range. 
(We associate a systematic error of 25 per cent with the extrapolated 
central value.)

\subsection{The cluster mass model}

\begin{figure}
\vspace{2cm}
\hbox{
\hspace{0.0cm}\psfig{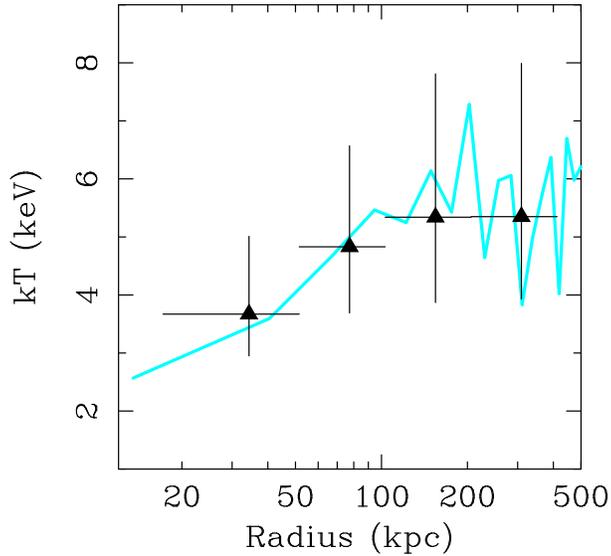}
}
\vspace{1cm}
\caption{The intrinsic temperature profile determined from the spectral 
deprojection analysis (Section 4.4; filled triangles) overlaid on the model 
temperature profile determined using the image deprojection code and 
best-fit NFW mass model (grey curve). The median model profile determined from 
100 monte-carlo simulations is shown).} \label{fig:kt_deproj}
\vskip 1cm
\end{figure}

The observed surface brightness and deprojected temperature profiles 
(Figs. \ref{fig:surbri}, \ref{fig:kt_deproj}) may together be 
used to measure the X-ray gas mass and total mass profiles 
in the cluster. A variety of simple parameterizations for the 
total mass distribution were examined, including 
an isothermal sphere (Equation 4-125 of Binney \& 
Tremaine 1987) and a Navarro, Frenk \& White (1997; hereafter NFW)
profile. In each case, the best-fit parameter values were 
determined using a simple iterative technique: given the observed 
surface brightness distribution and a particular parameterized mass model, 
the deprojection code is used to predict the temperature profile of the 
X-ray gas, which is then compared with the observations. (We use the median 
model temperature 
profile determined from 100 Monte-Carlo simulations.) The parameters for the 
mass model are stepped through a regular grid of values to determine the 
best-fit value (minimum $\chi^2$) and confidence limits.  
Spherical symmetry and hydrostatic equilibrium are assumed throughout.

We find that a good match to the observations can be obtained using an 
isothermal mass model with a core radius, $r_{\rm c} = 60\pm20$ kpc and a 
velocity dispersion $\sigma = 770\pm50$ \kmps (90 per cent confidence 
limits). Alternatively, the temperature and surface brightness data can be 
well-modelled using an NFW mass model with a scale radius, 
$r_{\rm s} = 0.19^{+0.17}_{-0.08}$ Mpc and a concentration parameter, 
$c=11.0^{+4.9}_{-3.7}$. The normalization of the NFW mass profile may also 
be expressed in terms of an equivalent velocity dispersion, $\sigma =  
\sqrt{50} H_0 r_{\rm s} c = 740^{+190}_{-120}$\kmps. The deprojected 
temperature profile predicted by the best-fitting NFW mass model is shown 
overlaid on the spectral results in Fig. \ref{fig:kt_deproj}. 

Smail \etal (1997) measure a projected mass within a radius of 400 kpc
of the centre of the 3C295 cluster from a weak lensing analysis of Hubble 
Space Telescope images of $4.7 \pm0.8 \times 10^{14}$\Msun. This 
value is approximately twice as large as the projected mass within the same 
radius implied by the best-fitting NFW X-ray mass model of 
$2.1^{+0.8}_{-0.5} \times 10^{14}$\Msun. (We assume that the X-ray mass 
model extends to an outer radius of 2 Mpc in our calculation of the 
projected mass). The discrepancy between the lensing and X-ray results is 
probably due to the presence of significant substructure 
along the line of sight to the cluster core, which includes the presence of a 
foreground cluster at $z \sim 0.26$ (Dressler \& Gunn 1992). It remains
possible, however, that the mean redshift of the lensed sources might exceed 
the value of 0.83 assumed in the Smail \etal (1997) study (which would cause 
the lensing mass to be over-estimated) and/or that the temperature of the 
X-ray gas might rise significantly beyond the central 250 kpc radius. Our 
results and conclusions regarding the projected mass within the central 400 
kpc region of the cluster are consistent with those of Neumann (1999).

\subsection{The X-ray gas properties}

The results on the the electron density and cooling time of 
the X-ray gas from the deprojection analysis, using the 
best-fitting NFW mass model, are shown in Fig. \ref{fig:deproj}. 
Within a radius of 500 kpc, the deprojected electron density 
distribution can be approximated ($\chi^2=47.6$ for 34 degrees of 
freedom) by a $\beta$-model with a core radius $r_c =17.8\pm1.3$ kpc, $\beta =
0.515\pm0.013$, and a central density, $n_{\rm e}(0) = 0.16\pm0.01$ 
cm$^{-3}$ ($1\sigma$ errors), although we caution that the results 
on the central density are sensitive, at the level of a few tens of per cent, 
to the manner in  which the X-ray emission associated with the radio lobes 
and central AGN are excluded.

For an assumed Galactic column density of $1.3\times10^{20}$\apc, we measure a 
central cooling time ({\it i.e.} the mean cooling time within the central 
2 arcsec bin) of $t_{\rm cool}=2.1\pm0.3 \times 10^8$ yr, and 
a cooling radius (at which the cooling time first exceeds a Hubble time) of 
$r_{\rm cool}=145^{+38}_{-30}$ kpc. 

\begin{figure}
\vspace{2cm}
\hbox{
\hspace{0.0cm}\psfig{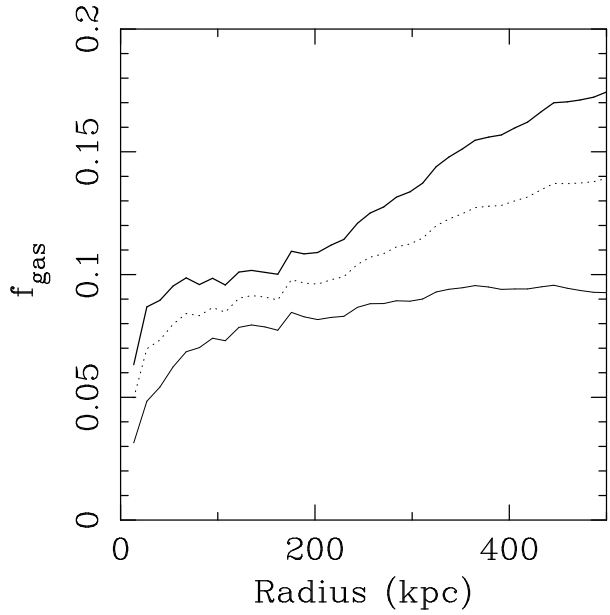}}
\vspace{1cm}
\caption{The X-ray gas-to-total mass ratio as a function of
radius determined from the deprojection analysis. From top 
to bottom, the three curves show the 90 per cent confidence upper 
limit, best-fit value (dotted curve), and 90 per cent confidence lower 
limit.}\label{fig:baryon}
\end{figure}

The X-ray gas-to-total mass ratio determined from the deprojection 
analysis is shown in  Fig. \ref{fig:baryon}. At $r = 500$ kpc (the 
outermost radius at which reliable temperature and density data are 
available) we measure 
$f_{\rm gas} =13.9^{+3.5}_{-4.6}$ per cent (90 per cent confidence limits).

\subsection{The mass deposition rate from the cooling flow}

\begin{figure}
\vspace{2cm}
\hbox{
\hspace{0cm}\psfig{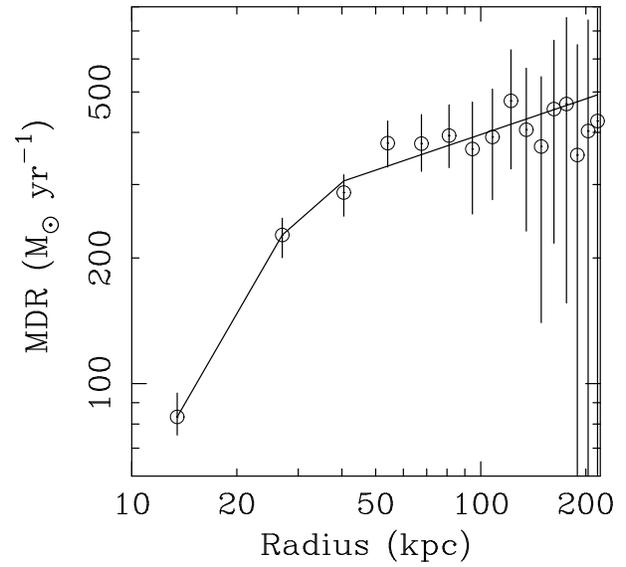}
}
\vspace{1cm}
\caption{The mass deposition profile from the cooling flow in 
3C295 determined from the deprojection analysis. A Galactic column density of $1.3 \times
10^{20}$\apc~is assumed. The best fitting broken power-law model is
overlaid.}\label{fig:break}
\end{figure}

The mass deposition profile from the cooling flow, determined from the 
image deprojection analysis, is shown in Fig. \ref{fig:break}. The 
mass deposition profile is a parameterization of the luminosity 
distribution in the cluster core, under the assumption that this emission is 
due to a steady-state cooling flow (for details of the calculation see 
\eg White \etal 1997). Since the presence of cooling gas will tend to 
enhance the X-ray luminosity of a cluster, the outermost radius at which 
cooling occurs may also be expected to be associated with a `break' in the 
X-ray surface brightness profile and, more evidently, the mass deposition 
profile determined from the deprojection code. We have fitted a broken 
power-law model to the mass deposition profile for 3C295 over the central 
200 kpc radius and find that the data exhibit a break at a radius of 
$32\pm5$ kpc, which is consistent with the break 
radius determined from the fit to the X-ray colour profile 
in Section 3.3. (The best-fit broken power-law model has been overlaid 
on the mass deposition profile in Fig. \ref{fig:break}.) 

The slopes of the mass deposition profile internal and
external to the break radius are $1.45\pm0.25$ and 
$0.28\pm0.13$, respectively ($1\sigma$ errors determined 
from a least-squares fit). Accounting only for absorption 
due to cold gas with the nominal Galactic column density, we measure an 
integrated mass deposition rate within the break radius determined from the 
deprojection analysis of $286^{+12}_{-59}$\Msunpyr. If we also account for 
the presence of an absorption edge with the best-fit properties determined 
with spectral model C2(v), the mass deposition rate within the break radius rises
to $316^{+13}_{-65}$\Msunpyr, in reasonable agreement with the 
independent spectral result of $281^{+85}_{-95}$\Msunpyr~(Table 3).

The Chandra result on the mass deposition rate is slightly lower than the 
value of $410^{+125}_{-130}$\Msunpyr~reported by White \etal (1997) based on 
Einstein Observatory High Resolution Imager (HRI) observations, and 
$400-900$\Msunpyr~reported by Neumann (1999) using ROSAT HRI data. 
However, both the White \etal (1997) and Neumann (1999) results were based on 
calculations which assumed a very old age for the cooling flow of 13 and 10 
Gyr, respectively. As we discuss below, the Chandra data suggest 
that the cooling flow in the 3C295 cluster is much younger than this, with 
an age of $\sim 1-2$ Gyr.

\subsection{The age of the cooling flow}

Allen \etal (2000) discuss a number of methods which may be used 
to estimate the ages of cooling flows. (Such ages are likely to relate 
to the time intervals since the central regions of clusters were 
last disrupted by major subcluster merger events.) Essentially, these 
methods identify the age of a 
cooling flow with the cooling time of the X-ray gas at the break 
radius measured in either the X-ray colour or deprojected mass 
deposition profiles.

\begin{figure*}
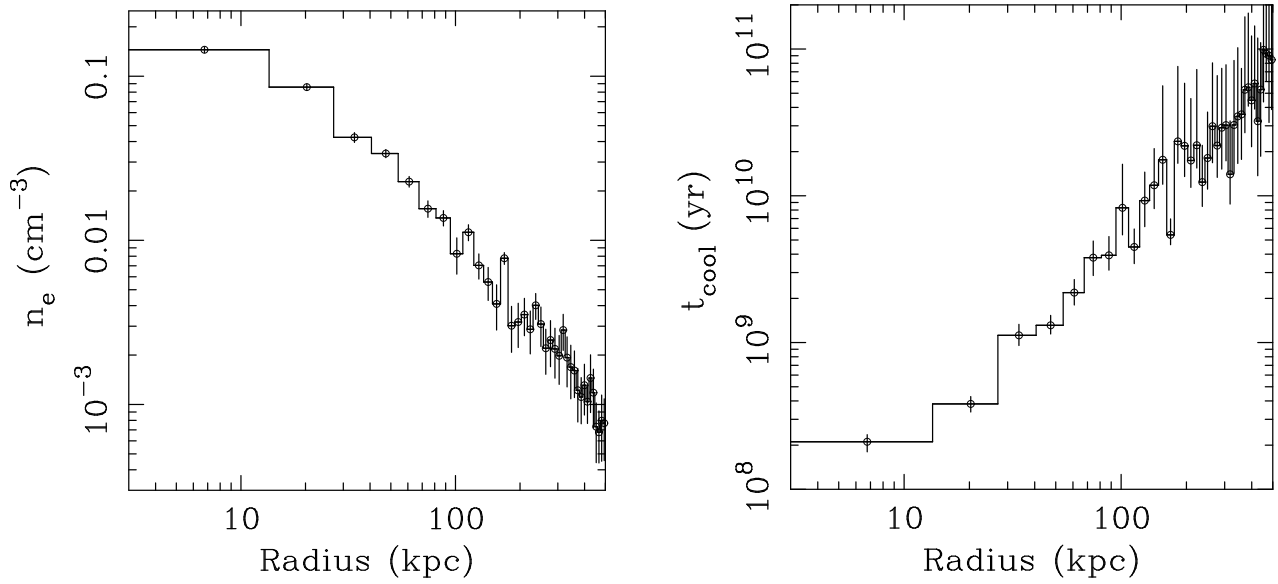

\vspace{2cm}
\hbox{
\hspace{0.0cm}\psfig{figure=ne_deproj_nfw_surbri3.ps,width=0.45\textwidth,angle=270}
\hspace{0.8cm}\psfig{figure=tcool_deproj_nfw_surbri3.ps,width=0.45\textwidth,angle=270}
}
\vspace{1cm}
\caption{The results on (a) the electron density and (b) the cooling time 
as a function of radius determined from the X-ray image deprojection analysis 
using the best-fit NFW mass model. A Galactic column density of $1.3 
\times 10^{20}$ \apc~is assumed. }\label{fig:deproj} 
\end{figure*}

Combining the results in Figs. \ref{fig:colour} and \ref{fig:deproj}, 
we see that cooling time of the X-ray gas at the break radius measured 
in the X-ray colour profile ($53^{+22}_{-19}$ kpc; Section 3.3) 
lies in the range $1.8^{+1.3}_{-0.9}$ Gyr. (The cooling time at the 
break radius is determined from a least-squares fit to the cooling time 
data in Fig. \ref{fig:deproj}(b) between radii of $20-80$ kpc using a 
power-law model.) If we instead identify the age of the cooling flow with 
the cooling time at the break radius in the mass deposition profile in 
Fig. \ref{fig:break} ($32\pm5$ kpc), we infer an age 
for the cooling flow of $0.8\pm0.2$ Gyr. In both cases we assume 
that no intrinsic absorption acts beyond the outer edge of the cooling 
flow, which is reasonable if the absorbing matter is accumulated by the flow.

Following Allen \& Fabian (1997), we may compare the observed
X-ray colour profile to the predictions from the deprojection
code for a steady-state cooling flow, assuming an age 
of 1.5 Gyr. This is shown in Fig. \ref{fig:colour}. The agreement 
between the observed and predicted values provides further support 
for the validity of the cooling-flow model in the central regions 
of the cluster.

In summary, we see that the two methods provide a consistent result on 
the age of the cooling flow in 3C295, with a preferred value of
$1-2$ Gyr. This value is less than the typical age determined for 
the sample of nearby cluster cooling flows studied by Allen \etal (2000), 
although implies a similar formation redshift. It is worth emphasising the 
remarkable capabilities of Chandra, which allow it to constrain the 
age of the cooling flow in a cluster at a redshift of $z=0.46$, using only a 
17ks exposure. The application of similar techniques to Chandra observations 
of a large sample of clusters should become possible over
 the next few years. This will provide significant insight into the
 evolution of cluster cooling flows and an important new data set for 
 comparison with numerical simulations of cluster formation.

\section{Radio properties}

\subsection{The age of the radio source}

Based on synchrotron aging arguments and ram pressure confinement
of the hot spots, Perley \& Taylor (1991) argue that the central radio
source in 3C295 has an age of 10$^{6}$ years. The age of the 
radio source is therefore $\sim 1000$ times less than the age we derive for the 
cooling flow. The total luminosity of 3C295 in the radio band is $\sim 
10^{45}$ \ergps. Assuming a radiative efficiency of 10 per cent, the 
`waste' energy supplied by the radio source into its environs is 
then on the order of 10$^{46}$ \ergps, which exceeds the 
absorption-corrected luminosity of the cluster in X-rays. Thus, if the 
central radio source were intermittently active on a time scale of $\sim 
10^8$ years, it could have a significant heating effect on the central 
X-ray gas. However, the cooling flow in 3C295 appears to have remained 
relatively undisturbed for a period of $1-2$ Gyr. 

\subsection{Faraday rotation measure and magnetic fields in the cluster gas}

The compact, strong hot-spots in 3C295 have minimum pressures of up
to 7.3 $\times$ 10$^{-8}$ dyn cm$^{-2}$.  Our deprojection analysis of
the Chandra observations leads to a central pressure approximately three times 
higher than that estimated from Einstein Observatory observations by Henry \& Henriksen (1986). 
In spite of the upwardly revised estimate of the central X-ray pressure, the 
radio pressures of the hot spots still exceed the thermal gas pressure at the
same radius by a factor of $\sim 100$ and ram pressure confinement of the hot spots is required. The radio
lobes, however, have minimum pressures of $10^{-9}$ dyn cm$^{-2}$ and may
be in equipartition with the X-ray emitting gas.  In fact, the high
brightness of the lobes in 3C295 may be a consequence of its
high-pressure environment.  Other examples of luminous radio galaxies
embedded in cooling flow clusters are 3C84 in the Perseus cluster,
Cygnus-A and Hydra A. Overall there is a trend for cD galaxies in 
dense X-ray emitting clusters to host moderately strong radio 
galaxies (\eg Ledlow \& Owen 1995), but most are less luminous than 3C295.

The radio galaxy 3C295 has been known for some time to possess very high
Faraday rotation measures (RM; Perley \& Taylor 1991), reaching 9000 rad
m$^{-2}$, whereas typical background radio sources along this line of
sight through our Galaxy have RMs of just 20 rad m$^{-2}$
(Simard-Normandin, Kronberg, \& Button 1981).  The most
likely origin for the high RMs in 3C295 is a cluster magnetic field.  To
date, all radio galaxies found at the center of dense X-ray emitting
clusters exhibit high RMs, with the RM correlating with the central
density excess or cooling flow rate (Taylor, Allen \&
Fabian 1999; Taylor, Barton \& Ge 1994). 

Faraday rotation of the radio emission from 3C295 will be produced 
if magnetic fields are intermixed with the hot X-ray emitting gas
that surrounds the radio galaxy.
The polarization angle obeys a $\lambda^2$ law and the relation 
between the RM, the gas density $n_e$, and the magnetic field
along the line of sight, $B_\|$, is given by

\begin{equation}
\label{RM}
RM = 812\int\limits_0^L n_{\rm e} B_{\|} {\rm d}l ~{\rm radians~m}^{-2}~,
\end{equation} 
where $n_e$ is the electron density in cm$^{-3}$, $B_{\|}$
is in $\mu$Gauss, and $l$ is in kpc.  On average, the magnetic field, $B$,
will be larger than the parallel component, $B_{\|}$, by $\sqrt{3}$.     

If the magnetic field is tangled with cells of uniform size, same
strength, and random orientation, then the observed RM along any given
line of sight will be generated by a random walk process.  The
expected distribution is Gaussian with a zero mean, and the dispersion
is related to the number of cells along the line of sight.  The source
will also depolarize if the cell size is smaller than the instrumental
resolution.  

\begin{figure}
\vspace{2cm}
\hbox{
\hspace{0cm}\psfig{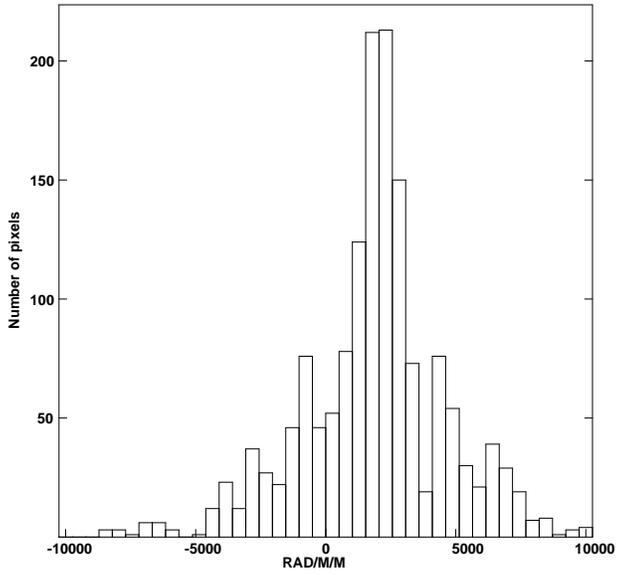}
}
\vspace{1cm}
\caption{Histogram of the rotation measure for all significant pixels
in 3C295 derived from the data of Perley \& Taylor (1991).}
\label{fig:rmhist}
\end{figure}

Perley \& Taylor (1991) measured the RM distribution in 3C295 and, 
assuming a constant density and magnetic field strength over 
a 100 kpc path, derived a field strength of 30 $\mu$G.  Using the
density profile derived in Section 5.2, and an analysis of the RM dispersion,
we can provide a refined estimate for the magnetic field strength.
From the patchy appearance of the Rotation Measure image, Perley \&
Taylor (1991) estimate a coherence length, $l$, of 0.5 arcsec (3.4 kpc).
Using a spherical $\beta$-model for the density
distribution, the expected dispersion in the RM for a source at the
same distance from us as the cluster center has been evaluated by
Felten (1996) to be 

\begin{equation}
\sigma_{RM} = {{441 B_\| n_0 r_{c}^{1/2} l^{1/2}} \over
(1 + r^2/r_c^2)^{(6\beta-1)/4}} \sqrt{\Gamma(3\beta -
  0.5)\over {\Gamma(3\beta)}}, 
\end{equation}

\noindent
where $\Gamma$ is the Gamma function. Given the RM dispersion in
3C295 (see Fig.\ref{fig:rmhist}), computed to be 2900 rad m$^{-2}$, we
estimate $B_\| = 7.2 \mu$G, implying $B \sim 12\mu$G 
(after scaling by $\sqrt{3}$).  The greatest uncertainty in this 
calculation stems from the knowledge of the cell size.  If the cell
size is in the range 10--1 kpc then the corresponding magnetic field
strength, $B$, is 7--20 $\mu$G.  These values
are quite similar to the field strengths of 5--10 $\mu$G found in A119
(Feretti et al.\ 1999), and 7--14 $\mu$G in the 3C129 cluster (Taylor 
\etal 2001) by a similar analysis.

\section{Discussion} 

\subsection{X-ray absorption by dust grains}

The absorption edge detected in the central regions of 3C295 
is presumably due to oxygen, the element with the largest absorption 
cross section in this region of the X-ray spectrum. If we assume that the 
oxygen was initially contained in gas with approximately solar metallicity, 
and that this material lay in a uniform screen in front of the X-ray 
emitting region, we may estimate the corresponding gas mass as 

\begin{equation}
M_{\rm abs} \sim 3 \times 10^7 r_{\rm abs}^2 \Delta N_{\rm H} \Msun,
\label{swa_eq3}
\end{equation}

\noindent where $r_{\rm abs}$ is the radial extent of the absorber in 
kpc and $\Delta N_{\rm H}$ is the equivalent hydrogen column density 
in units of $10^{21}$ \apc. For $r_{\rm abs} = 50^{+30}_{-20}$ kpc and 
$\Delta N_{\rm H} = 2.7^{+2.3}_{-1.0} \times 10^{21}$\apc~(Section 4.3.2), 
we obtain $M_{\rm abs} = 2.0^{+7.6}_{-1.5}\times 10^{11}$ \Msun~(although 
Allen \& Fabian 1997 and Wise \& Sarazin 2001 argue that for a geometry 
in which the absorbing material is distributed throughout the X-ray 
emitting region, the true mass may be a few times higher.)

This gas mass may be compared to the mass expected to have been accumulated 
by the cooling flow within the same radius over its lifetime 
($\sim 1-2$ Gyr; Section 5.4). If from time $t=0$ to $t=2 \times 10^9$ yr
the integrated mass deposition rate 
increases approximately linearly with
time, then after $t=2 \times 10^9$ yr, 
the accumulated mass is ${\dot M}t/2 \sim 4 \times 10^{11}$
\Msun, in reasonable agreement with the inferred mass of absorbing gas.
This suggests that the metals responsible for the 
X-ray absorption could initially have been associated with the cooled
gas deposited by the cooling flow.

A cold gaseous absorption model is not favoured by the Chandra 
data. The absence of significant absorption below the observed edge energy 
suggests that relatively little neutral hydrogen and helium are present and 
that the oxygen is therefore likely either to have been depleted onto dust 
grains or be in a warm, partially ionized state. (Intrinsic column 
densities of solar metallicity cold gas $\approxgt 2 \times 10^{20}$ \apc, 
included in 
addition to the absorption edge, can be excluded at 90 per cent confidence 
by the Chandra data.) 

The possibility that the X-ray absorption commonly detected in cooling 
flow clusters may be primarily due to dust grains has been investigated 
by several authors (Voit \& Donahue 1995; Arnaud \& Mushotzky 1998; 
Allen 2000; see also Fabian, Johnstone \& Daines 1994). 
In the case of silicate dust grains, the OI\,K edge at a rest-frame energy, 
$E\sim 0.54$keV is expected to be strong, with 
relatively little absorption at lower energies.
Arnaud \& Mushotzky (1998) presented Broad Band X-ray 
Telescope observations of the nearby Perseus cluster and
showed that the $0.35-7.0$ keV data require significant excess absorption 
which is better-explained by a simple oxygen 
absorption edge (which they associate with dust) than by a cold, gaseous 
absorption model. The introduction of an OI\,K edge also typically provides 
at least as good a description as a cold gaseous absorber for the 
intrinsic absorption detected in $0.6-10.0$ keV ASCA spectra for 
cluster cooling flows (\eg Allen \etal 2000 and references therein; in the 
ASCA band, the effects of a photoelectric absorption edge at 0.54 keV and a 
cold, gaseous absorber with approximately solar metallicity are essentially 
indistinguishable).

Optical and UV spectroscopy of the central $5-20$ kpc regions of 
cluster cooling flows often reveals large amounts of ongoing star 
formation (\eg Johnstone, Fabian \& Nulsen 1987; Allen 1995; 
Cardiel \etal 1995, 1998; McNamara \etal 1996; Voit \& Donahue 1997, 
Crawford \etal 1999; McNamara 1999) and significant intrinsic 
reddening due to dust (Hu 1992; Allen \etal 1995; Crawford \etal 1999). 
Infrared (\eg Allen \etal 2000; Hansen \etal 2000) and sub-mm 
(Edge \etal 1999) observations also require the presence of 
significant dust masses in the core regions of at least a few nearby 
and/or exceptionally massive cooling flows.  
In general, the data are consistent with dust being a ubiquitous
feature of the central regions of cluster cooling flows. 

The lifetime of grains of radius $a\mu$m to sputtering in hot gas of 
density $n$ is $\sim 2\times 10^6 a/n\yr$ (Draine \& Salpeter 1979). 
In the central 50 kpc of 3C295, $n \sim 5 \times 10^{-2}$cm$^{-3}$, so that 
for large dust grains with $a\sim 10\mu$m (which are inferred to have 
formed and carry most of the iron in \eg the expanding remnant of SN1987A; 
Colgan \etal 1984) the sputtering timescale is $\sim 4 \times 10^8$ yr. 
However, magnetic shielding in cooled clouds deposited from the cooling 
flow may allow dust grains to survive for significantly longer 
(B. Chandran, private communication). 

\subsection{The feasibility of a warm absorber}

The possible presence of warm, partially-ionized, spatially-extended 
absorbing gas associated with cooling flows in nearby galaxies and galaxy 
groups has recently been explored by Buote (2000a,b) using 
ROSAT Position Sensitive Proportional Counter (PSPC) data. 
Buote (2000a,b) shows that the introduction of a simple photoelectric 
absorption edge at $E \sim 0.54$ keV in the modeling of the PSPC 
spectra for the central regions of such systems leads to  
significant improvements in the goodness of fit, in an analogous manner to 
the results reported here. The explanation for the origin of the 
absorption favoured by Buote (2000a,b) is that it is due warm 
($T \sim$ a few $10^5$K), partially ionized gas. However, the existence 
of such material with masses, in some cases, exceeding the X-ray emitting 
gas mass within the same radii by factors of $1-10$ is difficult to 
understand in terms of its thermal stability and pressure support.

In the case of 3C295, a warm, ionized absorber is 
consistent with the observed edge energy, which allows for any 
ionization state of oxygen from OI-OV. However, the cooling 
rate of such material would be extremely rapid with a cooling 
time $\sim 100$ yr and an implied luminosity 
$L_{\rm abs} \sim 3 \times 10^{47} {(M_{\rm abs}/10^{12}M_{\odot}}){(T/10^5 K)}^{-3/2}$ \ergps, 
occurring mainly in optical/UV line emission (White \etal 1991; see also 
Loewenstein \& Fabian 1990, Daines, Fabian \& Thomas 1994; Pistinner 
\& Sarazin 1994). For $M_{\rm abs} \sim 2 \times 10^{11}$\Msun~and 
$T \sim 10^5$ K, $L_{\rm abs} \sim 6 \times 10^{46}$ \ergps, which 
significantly exceeds the total X-ray luminosity of the cluster. 
Although detailed UV observations of 3C295 are not currently available 
to unambiguously test this hypothesis, it is clear that an enormous 
heat source would be required to maintain the gas at such temperatures.

One possible mechanism for heating the absorbing matter without 
significantly heating the surrounding hot, X-ray emitting gas 
(which has the larger volume filling factor and which shows no obvious 
evidence for excess heating) is photoionization by 
UV/X-ray radiation from the central AGN. 
However, detailed calculations carried out by us using the Cloudy 
code (Ferland 1996; version 94.00) show that one cannot match
the required absorbing column density in ionized Oxygen without 
greatly exceeding the observed optical line luminosities 
(Baum \& Heckman 1989; Lawrence \etal 1996), other than with 
highly contrived models (see Appendix A for details).

A further difficulty with the warm absorber scenario in the case of 
3C295 is in preventing the warm, absorbing gas clouds from falling rapidly 
into to the cluster centre. To avoid falling inwards, a warm cloud must have 
sufficient size such that the drag due to the hot gas is negligible.  
Minimally, this requires that their terminal velocities exceed the 
Kepler speed {\it i.e.}

\begin{equation}
{\rho_w \over \rho_h} {r \over R} > 1, 
\end{equation}

\noindent where $\rho_w$ is the density of the warm gas, $\rho_h$ the density
of the hot gas, $r$ the radius of the warm cloud and $R$ the distance to the
cluster centre. Even then, the decay time of a cloud's orbit due to
drag from the hot gas can be short;

\begin{equation}
t_{\rm decay} \sim {\rho_w \over \rho_h} {r \over R} {R \over v_K},
\end{equation}

\noindent where $v_K$ is the Kepler speed, which should exceed the age of the
cooling flow. For a Kepler speed of $\sim 1000$ \kmps, an age of $2 
\times 10^9$ yr and $R = 50$ kpc, this requires 
 
\begin{equation}
{\rho_w \over \rho_h} {r \over R} > 40, 
\end{equation}

\noindent which can be translated into a limit on the mass of a warm
cloud of

\begin{equation}
{4\pi\over 3} \rho_w r^3 > {4\pi\over 3} \rho_h R^3 
40^3 \left( \rho_h \over \rho_w \right)^2
\end{equation}

\noindent ie the mass of one warm cloud must exceed the mass of all the hot gas
(roughly the total mass of cooled gas) unless $\rho_w > 250 \rho_h$.
The last condition is barely satisfied if the warm gas is in pressure
equilibrium and its temperature is $2 \times 10^5$ K.  
A further issue is then whether thermally unstable clouds, 
which almost certainly fail the first condition above when they are 
first formed, can aggregate to form such a large cloud on a 
sufficiently fast timescale. (We note that magnetic fields in the warm 
gas clouds, if sufficiently strong and stable, may contribute to the pressure 
support of the clouds and slightly reduce the cooling rate, although the
cooling will still be very rapid.)
It therefore appears that a warm, partially ionized absorber with 
$T\sim 10^5$ K cannot provide a satisfactory explanation for the 
X-ray absorption detected in the central regions of 3C295.

\subsection{On cooling flow models with a low temperature cut-off}

Since this paper was submitted, a series of preprints based on the analysis 
of XMM-Newton data for the cooling flow clusters Abell 1795 (Tamura \etal 
2001), Abell 1835 (Peterson \etal 2001) and Sersic 159-03 (Kaastra \etal 
2001) have appeared. These works argue that the XMM-Newton data constrain 
the mass deposition rates from the cooling flows in these clusters, 
measured using a constant-pressure model, to be significantly 
less than the values determined from previous image deprojection studies 
which assumed ages for the cooling flows of $\sim 13$ Gyr, and studies of 
the integrated cluster spectra observed with ASCA. A model favoured by 
these authors is a modified cooling flow model with a low temperature 
cut-off, $kT_{\rm min}$, below which the gas does not cool. These 
authors measure values for $kT_{\rm min}$ of $\sim 2.4$ keV (Abell 1795; 
Tamura \etal 2001), $\sim 2.7$ keV (Abell 1835; Peterson \etal 2001) and 
$\sim 1.4$ keV (Sersic 159-03; Kaastra \etal 2001), respectively.

We have explored whether a cooling flow model with a low temperature cut-off 
can also provide a good description of the Chandra data for the 3C 295 
cluster. Fitting the $0.3-7.0$ keV spectrum for the central 100 kpc radius 
with spectral models C1(v) and C2(v) including such a cut-off, we 
measure $kT_{\rm min} < 1.0$ keV (model C1v) and $kT_{\rm min} < 0.7$ keV 
(model C2v) at 90 per cent confidence, respectively (other parameters in 
good agreement with the values listed in Table 3). Over the restricted 
$0.5-7.0$ keV range, we measure $kT_{\rm min} < 1.2$ keV (model C1v) and 
$kT_{\rm min} < 0.8$ keV (model C2v). Thus, the Chandra data for the 
central regions of the 3C 295 cluster do not require the introduction of a 
low temperature cut-off in the cooling flow model and constrain the 
temperature of any such cut-off to be $kT_{\rm min} \approxlt 1$ keV.

Finally, we note that when comparing results on the properties of cooling 
flows determined from spatially-resolved spectroscopy with XMM-Newton and 
Chandra with the findings from previous X-ray missions, it is important to 
consider the strong assumptions involved in the earlier work. Firstly, 
the typical ages of cooling flows are likely to be significantly less than 
the canonical, maximum value of 13 Gyr assumed in most previous X-ray 
imaging studies (Section 5.4; see also \eg Schmidt, Allen \& Fabian 2001). 
Thus, the true mass deposition rates from the cooling flows are also likely 
be lower than the previously-published maximal values. Secondly, as discussed 
by Allen, Ettori \& Fabian (2001), the presence of strong ambient temperature 
gradients in the centres of 
(at least some) cooling-flow clusters are likely to have caused the mass 
deposition rates inferred from previous spectral studies based on the 
analysis of integrated cluster spectra (which could not resolve spatial
variations in the ambient cluster temperature), to have overestimated 
(\eg by a factor $\sim 3$ in the case of the massive cluster 
Abell 2390) the true mass deposition rates from the cooling flows.

\section{Conclusions}

The main conclusions that can be drawn from this work may be 
summarized as follows:
\vskip 0.2cm

(i) The 17ks Chandra ACIS-S observation of the cluster of galaxies 
surrounding the powerful radio source 3C295 ($z=0.46$) provide firm 
constraints on the properties of the intracluster gas. Between radii of 
$50-500$ kpc, the X-ray gas appears approximately isothermal with a mean, 
emission-weighted temperature $kT= 5.3^{+1.2}_{-0.7}$ keV (90 per cent 
confidence limits). Within the central 50 kpc radius, this 
value drops to $kT \sim 3.7$ keV. The mean metallicity across 
the central 250 kpc radius is $Z \sim 0.6$ solar. 

\vskip 0.2cm
(ii) The spectral and imaging Chandra data indicate the presence of a 
cooling flow within the central 50kpc radius of the cluster, 
with a mass deposition rate of $\sim 280$ \Msunpyr. We estimate an 
age for the cooling flow of $1-2$ Gyr, which is $\sim 1000$ times older 
than the central radio source. The presence of the cooling flow suggests 
that the radio source has not significantly heated the surrounding X-ray gas. 
On larger scales, the X-ray gas appears regular and relaxed.

\vskip 0.2cm
(iii) The Chandra spectrum for the central 50 kpc region associated with 
the cooling flow exhibits an edge-like absorption feature which can be 
interpreted as being due to oxygen-rich dust grains. The implied mass in 
metals seen in absorption may have been accumulated by the cooling flow 
over its lifetime.

\vskip 0.2cm
(iv) Combining the X-ray gas density profile measured with Chandra with 
radio determinations of the Faraday rotation measure in 3C295, we 
estimate the magnetic field strength in the region of the cluster 
core to be $B \sim 12 ~\mu$G.

\section*{Acknowledgements}

SWA and PEJN acknowledge the hospitality of the Harvard-Smithsonian
Center for Astrophysics. We thank Massimo Cappi, Maxim Markevitch 
and Robert Schmidt for helpful discussions. We also thank the 
anonymous referee for suggestions which helped to improve the clarity of 
the paper. SWA and ACF thank the Royal Society for support.

\clearpage

\clearpage

\clearpage

\clearpage

\clearpage

\begin{figure}
\vspace{2cm}
\hbox{
\hspace{0cm}\psfig{figure=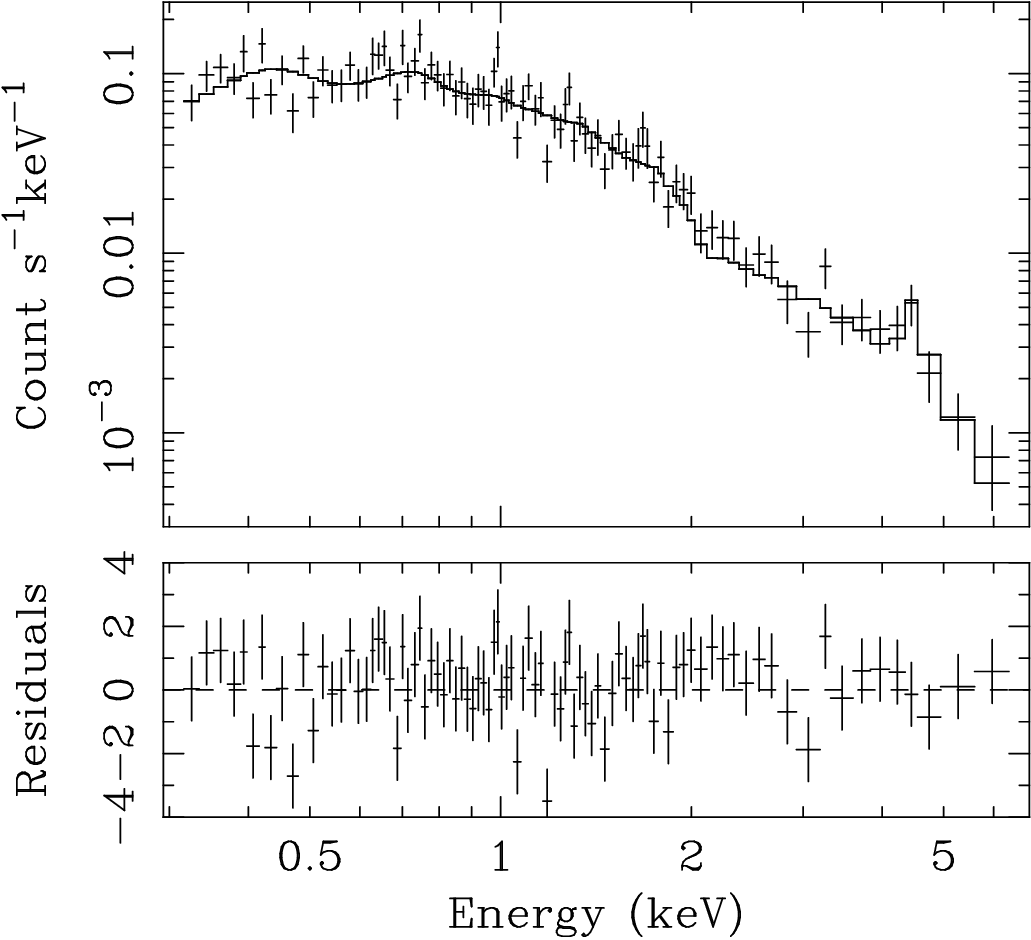,width=0.45 \textwidth,angle=270}
}
\vspace{1cm}
\caption{Chandra spectrum for the $50-250$ kpc
annulus with the best-fit single-temperature
plasma model (model B) overlaid. Other details as in Fig. \ref{fig:spec_core}.
}\label{fig:spec_ann1}
\end{figure}

\begin{figure}
\vspace{2cm}
\hbox{
\hspace{0cm}\psfig{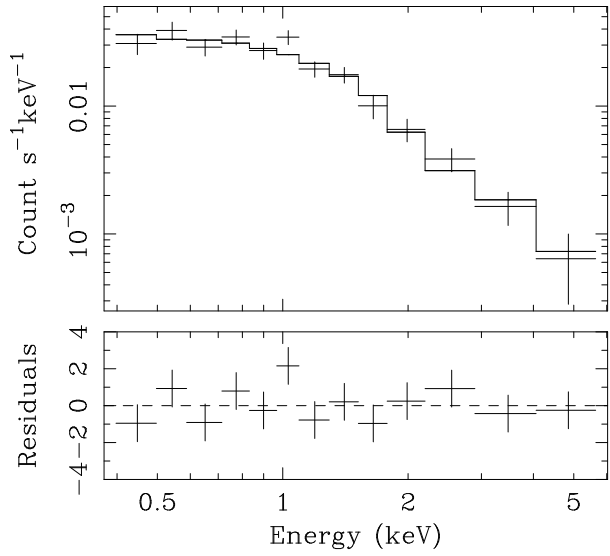}
}
\vspace{1cm}
\caption{Chandra spectrum for the $250-500$ kpc
annulus with the best-fit single-temperature
plasma model (model B) overlaid.  Other details as in Fig. \ref{fig:spec_core}.
}\label{fig:spec_ann1pt5}
\end{figure}

\clearpage

\clearpage

\clearpage

\clearpage

\clearpage

\clearpage


\clearpage

\begin{table*}
\vskip 0.2truein
\begin{center}
\caption{The best-fit parameter values and 90 per cent  ($\Delta \chi^2 = 2.71$) 
confidence limits determined from the spectral analysis of the Chandra data 
using the single-temperature emission models (models A and B). Temperatures 
($kT$) are quoted in keV, metallicities ($Z$) as a fraction of the solar 
value (Anders \& Grevesse 1989) and X-ray absorbing column densities 
($N_{\rm H}$) in units of of $10^{20}$ atom cm$^{-2}$.}
\vskip 0.2truein
\begin{tabular}{ c c c c c c c c c }
\hline
\multicolumn{1}{c}{} &
\multicolumn{1}{c}{} &
\multicolumn{3}{c}{ENERGY RANGE $0.5-7.0$ keV} &
\multicolumn{1}{c}{} &
\multicolumn{3}{l}{ENERGY RANGE $0.3-7.0$ keV} \\
&&&&&&&&\\
             & ~ & $0-50$kpc  &  $50-250$kpc & $250-500$kpc & ~ & $0-50$kpc  &  $50-250$kpc & $250-500$kpc  \\
&&&&&&&&\\
MODEL A&&&&&&&& \\
$kT$         & ~ & $3.57^{+0.58}_{-0.47}$ & $5.02^{+0.80}_{-0.61}$ & $6.4^{+4.3}_{-2.2}$  &~& $3.72^{+0.65}_{-0.49}$ & $5.22^{+0.85}_{-0.65}$ & $6.57^{+3.92}_{-2.08}$    \\
$Z$          & ~ & $0.59^{+0.39}_{-0.32}$ & $0.60^{+0.36}_{-0.31}$ & $<1.1$               &~& $0.59^{+0.40}_{-0.34}$ & $0.60^{+0.36}_{-0.32}$ & $<0.90$                   \\
$\chi^2$/DOF & ~ & 35.2/40                & 87.3/76                & 8.9/8                &~& 48.3/48                & 110.6/88               & 10.7/10                   \\
&&&&&&&& \\ 
MODEL B&&&&&&&&\\                                                                         
$kT$         & ~ & $3.79^{+0.66}_{-0.67}$ & $5.44^{+0.95}_{-0.88}$ & $4.7^{+6.2}_{-2.0}$  &~& $3.76^{+0.84}_{-0.59}$ & $5.27^{+0.95}_{-0.88}$ & $5.43^{+5.46}_{-2.12}$    \\
$Z$          & ~ & $0.59^{+0.41}_{-0.34}$ & $0.59^{+0.32}_{-0.37}$ & $<2.1$               &~& $0.60^{+0.42}_{-0.34}$ & $0.60^{+0.32}_{-0.37}$ & $<0.93$                   \\
$N_{\rm H}$  & ~ & $<3.1$                 & $<2.1$                 & $<16.9$              &~& $1.19^{+1.51}_{-1.19}$ & $1.23^{+1.13}_{-1.04}$ & $2.89^{+4.06}_{-2.89}$    \\
$\chi^2$/DOF & ~ & 34.7/39                & 86.0/75                & 8.4/7                &~& 48.3/47                & 110.6/87   & 10.2/9    \\      
&&&&&&&&  \\    
\hline 
\end{tabular}
\end{center}  
\parbox {7in}
{} 
\end{table*}

\begin{table*}
\vskip 0.5truein
\begin{center}
\caption{The best-fit parameter values and 90 per cent confidence limits from the 
multiphase spectral analysis of the central 50 kpc region using the data in 
the $0.5-7.0$ keV band. The mass deposition rate from the cooling flow
(${\dot M}$) is given in units of \Msunpyr. Other details as in Table 1.}
\vskip 0.2truein
\begin{tabular}{ c c c c c }
\hline                                                              
\multicolumn{1}{c}{} &
\multicolumn{1}{c}{} &
\multicolumn{1}{c}{} &
\multicolumn{2}{c}{EMISSION MODEL} \\
ABSORPTION MODEL &             & ~ &  C1  &  C2  \\
&&&& \\
&$kT_1$                        & ~ & $4.30^{+5.62}_{-0.93}$ & $4.18^{+3.10}_{-0.82}$ \\       
CASE (i)            &$Z$       & ~ & $0.60^{+0.44}_{-0.35}$ & $0.60^{+0.44}_{-0.35}$ \\     
GALACTIC ABSORPTION 
&${\dot M}$                    & ~ & $78^{+86}_{-75}$       & $77^{+86}_{-74}$       \\                         
&$\chi^2$/DOF                  & ~ & 32.3/39                & 32.3/39                \\                     
&&&& \\		      		                                                                                  

&$kT_1$                        & ~ & $6.99^{+3.42}_{-3.42}$ & $4.66^{+2.35}_{-1.35}$ \\      
CASE (ii) &$Z$                 & ~ & $0.60^{+0.42}_{-0.32}$ & $0.58^{+0.45}_{-0.30}$ \\     
VARIABLE  ABSORPTION
                  &${\dot M}$  & ~ & $317^{+172}_{-208}$    & $333^{+182}_{-218}$    \\                         
BY COLD GAS (z=0)
&$N_{\rm H}$                   & ~ & $8.29^{+4.86}_{-4.77}$ & $8.51^{+4.94}_{-5.43}$ \\     
&$\chi^2$/DOF                  & ~ & 28.3/38                & 28.1/38                \\            
&&&& \\					 
\hline    
\end{tabular}
\end{center}
\parbox {7in}
{}
\end{table*}

\begin{table*}
\vskip 0.5truein
\begin{center}
\caption{The best-fit parameter values and 90 per cent confidence limits 
from the multiphase spectral analysis of the central 50 kpc region using 
the data from 
the full $0.3-7.0$ keV band. The mass deposition rate from the cooling flow
(${\dot M}$) is in units of \Msunpyr. Column densities ($N_{\rm H}$) and  
intrinsic column densities ($\Delta N_{\rm H}$) are in units of $10^{20}$ atom 
cm$^{-2}$. The absorption edge energy ($E_{\rm edge}$) is in units of keV.
Other details as in Table 1. For a complete description of the spectral models see Section 
4.2}
\vskip 0.2truein
\begin{tabular}{ c c c c c c }
\hline                                                              
\multicolumn{1}{c}{} &
\multicolumn{1}{c}{} &
\multicolumn{1}{c}{} &
\multicolumn{3}{c}{EMISSION MODEL} \\
\multicolumn{1}{c}{} &
\multicolumn{1}{c}{} &
\multicolumn{1}{c}{} &
\multicolumn{1}{c}{SINGLE-PHASE~} &
\multicolumn{2}{c}{MULTIPHASE} \\
ABSORPTION MODEL &             &  & A &  C1  &  C2  \\
&&&&& \\
&$kT_1$                  &  & $3.72^{+0.65}_{-0.49}$ &$3.99^{+1.67}_{-0.72}$ & $3.96^{+1.48}_{-0.69}$  \\
CASE (i)            &$Z$ &  & $0.59^{+0.40}_{-0.34}$ &$0.61^{+0.43}_{-0.35}$ & $0.61^{+0.43}_{-0.35}$  \\
GALACTIC ABSORPTION 
&${\dot M}$              &  & ---                    &$28^{+69}_{-28}$       & $28^{+69}_{-28}$        \\
&$\chi^2$/DOF            &  & 48.3/48                &47.9/47                & 47.9/47                 \\
&&&&& \\

&$kT_1$                  &  & $3.76^{+0.84}_{-0.59}$ &$3.98^{+1.80}_{-0.76}$ &$3.94^{+1.53}_{-0.72}$   \\
CASE (ii)           &$Z$ &  & $0.60^{+0.42}_{-0.34}$ &$0.61^{+0.43}_{-0.35}$ &$0.61^{+0.43}_{-0.35}$   \\
VARIABLE ABSORPTION 
&${\dot M}$              &  & ---                    &$34^{+105}_{-34}$      &$35^{+104}_{-35}$        \\
BY COLD GAS (z=0) 
&$N_{\rm H}$             &  & $1.19^{+1.51}_{-1.19}$ &$1.56^{+2.09}_{-1.56}$ &$1.57^{+2.09}_{-1.57}$   \\
&$\chi^2$/DOF            &  & 48.3/47                &47.9/46                &47.9/46                  \\
&&&&& \\

&$kT_1$                  &  & $3.61^{+0.71}_{-0.54}$ &$4.06^{+2.05}_{-0.77}$  & $1.97^{+0.85}_{-0.38}$  \\
CASE (iii)           &$Z$&  & $0.57^{+0.40}_{-0.32}$ &$0.57^{+0.40}_{-0.31}$  & $0.37^{+0.27}_{-0.15}$  \\
INTRINSIC ABSORPTION   
      &${\dot M}$        &  & ---                    &$120^{+107}_{-108}$     & $1796^{+2067}_{-1375}$    \\
BY COLD GAS ($z=z_{\rm clus}$)
&$\Delta N_{\rm H}$      &  & $0.84^{+2.55}_{-0.84}$ &$18.0^{+52.7}_{-8.6}$   & $584.3^{+310.2}_{-188.8}$ \\
&$\chi^2$/DOF            &  & 48.0/47                &45.1/46                 & 40.5/46                   \\

&&&&& \\
&$kT_1$                  &  & $2.15^{+0.73}_{-0.45}$ &$2.32^{+0.80}_{-0.46}$   & $2.04^{+0.78}_{-0.45}$    \\
CASE (iv) &$Z$           &  & $0.41^{+0.25}_{-0.17}$ &$0.45^{+0.29}_{-0.18}$   & $0.38^{+0.26}_{-0.16}$    \\
PARTIAL COVERING 
           &${\dot M}$   &  & ---                    &$3198^{+6289}_{-2604}$   & $1652^{+2212}_{-1237}$    \\
BY COLD GAS ($z=z_{\rm clus}$) 
&$\Delta N_{\rm H}$      &  & $586.2^{+330.2}_{-227.6}$  &$780.0^{+449.0}_{-302.7}$& $589.4^{+306.6}_{-195.4}$ \\
&$f$                     &  & $0.67^{+0.15}_{-0.28}$ &$1.00^{+0.00}_{-0.91}$   & $1.00^{+0.00}_{-0.91}$    \\
&$\chi^2$/DOF            &  & 40.8/46                &41.4/45                  & 40.5/45                   \\
&&&&& \\

&$kT_1$                  &  & $3.37^{+0.57}_{-0.53}$ &$7.58^{+2.66}_{-3.40}$ & $5.20^{+1.74}_{-1.52}$   \\
CASE (v)    &$Z$         &  & $0.53^{+0.34}_{-0.30}$ &$0.48^{+0.36}_{-0.26}$ & $0.47^{+0.35}_{-0.26}$   \\
SIMPLE EDGE &${\dot M}$  &  & ---            &$272^{+78}_{-68}$      & $281^{+85}_{-95}$        \\
($z=z_{\rm clus}$) &$E_{\rm edge}$          & ~ & $0.61^{+0.05}_{-0.08}$ &$0.63^{+0.03}_{-0.05}$ & $0.63^{+0.03}_{-0.05}$   \\
&$\tau$                  &  & $0.38^{+0.39}_{-0.27}$ &$1.15^{+1.52}_{-0.44}$ & $1.14^{+0.97}_{-0.41}$   \\
&$\chi^2$/DOF            &  & 43.3/46                &31.0/45                & 30.8/45                  \\
&&&&& \\

\hline    
\end{tabular}
\end{center}
\parbox {7in}
{}
\end{table*}

\clearpage

\appendix

\section{Photoionization modeling and the plausibility of a warm absorber}

As mentioned in Section 7.2, one possible mechanism for heating  
relatively cool, X-ray absorbing gas clouds without 
significantly heating the surrounding hot, X-ray emitting gas 
(which has the larger volume filling factor and which shows no obvious 
evidence for excess heating) is photoionization by 
UV/X-ray radiation from the central AGN. We have used version 
94.00 of the Cloudy photoionization code (Ferland 1996) to
investigate this possibility. Our modeling uses the TABLE AGN command 
to specify the continuum shape (in a similar manner 
to Matthews and Ferland 1987),
a pressure of $\sim10^7\pccmK$ (which approximates that
in the X-ray emitting gas at a radius of 
$\sim 10\kpc$ from the nucleus), an abundance set
matching that used in the X-ray spectral fitting and a heavy element
abundance of 0.5 solar. The irradiated clouds were assumed to
be free of grains. We have explored a range of ionizing fluxes and
examined the resulting column densities in various oxygen ions and
the predicted luminosities in a number of optical/UV emission lines.

We initially ran two models corresponding to total 
photoionizing luminosities of
$3\times10^{45}\ergps$ (model 1) and $10^{46}\ergps$ (model 2). The
predicted column densities in various oxygen ions are shown in Table
A1. The column density in ionized oxygen is very small when compared with 
that inferred from the fits to the X-ray spectra with the absorption edge 
model ($N_O=2.3^{+2.0}_{-0.9} \times 10^{18}\pscm$). 
The Cloudy models were terminated arbitrarily when the electron
temperature fell to 1000K, where the oxygen is predominantly neutral.
In principle, there may be an arbitrary amount of gas colder than 
this in the Cloudy models, subject to the constraint that it is Jeans stable.
However, the X-ray data only allow absorption from a full ISM-like absorber 
with an equivalent neutral hydrogen column density
$\approxlt 4\times 10^{20}\pscm$ (for half solar metallicity) in addition to the 
absorption edge, which would
imply a maximum extra column density in oxygen of $1.7\times
10^{17}\pscm$. Adding this to the sum of the oxygen column densities in
all ionization states ($1.5\times10^{17}\pscm$ from the Cloudy models with
an ionizing luminosity of $10^{46}\ergps$) gives a total possible column
density of $3.2\times10^{17}\pscm$, which is still an order of magnitude short
of the required value. We also note that allowing the
models to continue to temperatures below 1000K causes a large build up
of radiation pressure in excess of the thermal pressure.

Table A2 shows the predicted luminosities in the $H\beta$,
[OII]$\lambda\lambda 3727,3729$, [OIII]$\lambda 5007$, Ly$\alpha$ and
CIV $\lambda\lambda1548,1552$ lines. There are no ultraviolet
spectral observations of 3C295, but the observed
[OIII]$\lambda 5007$ luminosity determined from narrow band imaging is 
$\sim 2.5 \times 10^{42}$ \ergps (Baum \& Heckman 1989). Scaling the optical 
line fluxes measured by Lawrence \etal (1996) to the total 
[OIII]$\lambda 5007$ luminosity also implies total 
[OII]$\lambda\lambda 3727,3729$ and $H\beta$ luminosities 
of $\sim 4.7 \times 10^{42}$ and $6.1 \times 10^{41}$ \ergps, respectively. 
We therefore see that the optical line luminosities predicted by 
photoionization models 1 and 2 exceed the observed values by $2-3$ orders of 
magnitude, ruling out such models. (Similar results can be obtained 
using the tabulated line emissivities of Pistinner \& Sarazin 1994.)

A cooling cloud could, in principle, be at a pressure much lower than that
inferred from the X-ray emitting gas if there were substantial 
magnetic pressure support, or if the cloud were to cool isochorically. 
To investigate how this might change the column densities in various 
ionization states of oxygen and the predicted optical emission line 
luminosities, we have
run a series of further photoionization calculations at lower
pressures. Constant pressure models with
$nT\sim 10^{4} \pccmK$ were found to be dominated by radiation
pressure in diffuse emission lines and are probably 
unstable. We have also examined 
a constant density model (model 3): setting the hydrogen number density
to $0.32\pccm$ and using an ionizing flux appropriate for a nuclear ionizing
luminosity of $3\times 10^{45}\ergps$, we found that the pressure ranged from 
$nT\sim 1.4\times10^5-3.6\times10^3\pccmK$. For such a model, we obtain a much
higher degree of ionization, with the column densities in
various ionization states of oxygen given in Table A1. There is a substantial
amount of oxygen in the four times ionized state, but even more in
the twice ionized state. The total oxygen column density in this
model is $ 8.8\times 10^{18}\pscm$, and could be higher if the model
were allow to proceed to cooler temperatures. Again, however, Table A2 shows
that very large $H\beta$, [OIII], Ly$\alpha$ and CIV line luminosities 
are expected, which greatly exceed the observed line fluxes at optical 
wavelengths, although the [OII] emission is somewhat reduced due to the 
smaller amount of singly ionized oxygen. Finally, to reduce the [OIII] 
emission, we have artificially truncated the ionized cloud
at a size of $2.3\times10^{20}\cm$ (model 4). This reduces the [OIII]
emission by more than two orders of magnitude, whilst still
maintaining a large ionized oxygen column density ($\sim 7\times
10^{17}\pscm$). A model intermediate between models 3 and 4 could probably 
produce a sufficient column density in ionized oxygen without over-predicting
the optical [OIII]$\lambda 5007$ emission, but would be extremely contrived 
and still produce strong CIV $\lambda\lambda1548,1552$ lines. HST 
observations of the CIV emission lines in 3C295 would 
provide a definitive test of such a model.

\begin{table*}
\vskip 0.2truein
\begin{center}
\caption{ \label{table:columns}
Column densities of oxygen in various ionization states from the Cloudy  
modeling. Column 2 lists the photoionizing luminosity in units of 
$10^{45}$\ergps. Columns $3-10$ list the column densities in each 
ionization state of oxygen in \apc. 
}
\vskip 0.2truein
\begin{tabular}{ccccccccccc}
\hline
\multicolumn{1}{c}{} &
\multicolumn{1}{c}{IONIZING} &
\multicolumn{9}{c}{OXYGEN COLUMN DENSITY} \\
\multicolumn{1}{c}{MODEL} &
\multicolumn{1}{c}{LUMINOSITY} &
\multicolumn{1}{c}{$O^0$} &
\multicolumn{1}{c}{$O^+$} &
\multicolumn{1}{c}{$O^{2+}$} &
\multicolumn{1}{c}{$O^{3+}$} &
\multicolumn{1}{c}{$O^{4+}$} &
\multicolumn{1}{c}{$O^{5+}$} &
\multicolumn{1}{c}{$O^{6+}$} &
\multicolumn{1}{c}{$O^{7+}$} &
\multicolumn{1}{c}{$O^{8+}$} \\
&&&&&&&&& \\
1  & 3.0  & 1.8E16 & 2.7E16 & 1.3E16 & 3.3E14 & 3.8E12 &  ---   &  ---   & ---    & ---  \\
2  & 10.0 & 1.9E16 & 4.2E16 & 8.0E16 & 6.9E15 & 2.4E14 & 1.5E12 &  ---   & ---    & --- \\
3  & 3.0  & 3.0E18 & 1.5E17 & 4.4E18 & 5.2E17 & 5.5E17 & 1.8E17 & 2.8E16 & 8.8E13 & --- \\
4  & 3.0  & 1.5E08 & 5.6E12 & 8.9E15 & 2.0E17 & 3.4E17 & 1.3E17 & 2.1E16 & 6.8E13 & --- \\
\hline
\end{tabular}
\renewcommand{\baselinestretch}{1.0}
\newline
\end{center}
\parbox {7in}
{}
\vskip 0.5 truein
\end{table*}

\begin{table*}
\vskip 0.2truein
\begin{center}
\caption{ \label{table:lums} 
The predicted optical/UV line luminosities (in \ergps) determined from the 
photoionization modeling using the Cloudy code. The models are the same as 
in Table ~\ref{table:columns}.}
\vskip 0.2truein
\begin{tabular}{cccccc} 
\hline
MODEL & H$\beta$ & [OIII] & [OII]  & Ly$\alpha$ & CIV \\
&&&&& \\
1  & $3.0\times 10^{43}$ & $1.4\times 10^{44}$ & $2.2\times 10^{44}$ & $8.3\times 10^{44}$ & $3.3\times 10^{41}$ \\
2  & $9.1\times 10^{44}$ & $9.3\times 10^{44}$ & $4.1\times 10^{44}$ & $2.4\times 10^{45}$ & $9.6\times 10^{42}$ \\
3  & $3.0\times 10^{43}$ & $4.3\times 10^{44}$ & $8.7\times 10^{42}$ & $7.5\times 10^{44}$ & $9.1\times 10^{43}$ \\ 
4  & $2.1\times 10^{43}$ & $2.1\times 10^{42}$ &        ---          & $7.0\times 10^{43}$ & $2.1\times 10^{43}$ \\
\hline
\end{tabular}
\renewcommand{\baselinestretch}{1.0}
\newline
\end{center}
\parbox {7in}
{}
\end{table*}

\end{document}